\documentclass[sigconf]{acmart}
\usepackage{multirow}
\usepackage{algorithm}
\usepackage[algo2e,ruled,vlined]{algorithm2e}
\usepackage{graphicx}
\usepackage{amsmath}
\usepackage{graphicx}
\usepackage{booktabs}
\usepackage{multirow}
\usepackage{chngpage}
\usepackage{bbm}
\usepackage{enumitem}
\usepackage{subcaption}
\usepackage{amsmath}

\newcommand{\cutsectionup}{\vspace*{-0.00in}}
\newcommand{\cutsubsectionup}{\vspace*{-0.00in}}
\newcommand{\cutparagraphup}{\vspace*{0.00in}}
\newcommand{\cutcationup}{\vspace*{-0.04in}}

\newcommand{\model}[1]{\textsf{\small{FedDefender}}}

\renewcommand{\eqref}[1]{\ref{#1}}

\AtBeginDocument{%
  \providecommand\BibTeX{{%
    \normalfont B\kern-0.5em{\scshape i\kern-0.25em b}\kern-0.8em\TeX}}}

\copyrightyear{2023}
\acmYear{2023}
\setcopyright{acmlicensed}
\acmConference[KDD '23] {Proceedings of the 29th ACM SIGKDD Conference on Knowledge Discovery and Data Mining}{August 6--10, 2023}{Long Beach, CA, USA.}
\acmBooktitle{Proceedings of the 29th ACM SIGKDD Conference on Knowledge Discovery and Data Mining (KDD '23), August 6--10, 2023, Long Beach, CA, USA}
\acmPrice{15.00}
\acmDOI{10.1145/3580305.3599346}
\acmISBN{979-8-4007-0103-0/23/08}

\begin{document}
\title{FedDefender: Client-Side Attack-Tolerant Federated Learning}
\settopmatter{authorsperrow=4, printacmref=true}

\author{Sungwon Park}
\authornote{Equal contribution to this work}
\orcid{0000-0002-6369-8130}
\affiliation{%
  \institution{KAIST}
  \city{Daejeon}
  \country{South Korea}
}
\email{psw0416@kaist.ac.kr}

\author{Sungwon Han}
\authornotemark[1]
\orcid{0000-0002-1129-760X}
\affiliation{%
  \institution{KAIST}
  \city{Daejeon}
  \country{South Korea}
}
\email{lion4151@kaist.ac.kr}

\author{Fangzhao Wu}
\orcid{0000-0001-9138-1272}
\affiliation{%
  \institution{Microsoft Research Asia}
  \city{Beijing}
  \country{China}
}
\email{wufangzhao@gmail.com}

\author{Sundong Kim}
\orcid{0000-0001-9687-2409}
\affiliation{%
 \institution{GIST}
 \city{Gwangju}
 \country{South Korea}
 }
 \email{sundong@gist.ac.kr}

\author{Bin Zhu}
\orcid{0000-0002-3571-7808}
\affiliation{%
  \institution{Microsoft Research Asia}
  \city{Beijing}
  \country{China}
}
\email{binzhu@microsoft.com}

\author{Xing Xie}
\orcid{0000-0002-8608-8482}
\affiliation{%
  \institution{Microsoft Research Asia}
  \city{Beijing}
  \country{China}
}
\email{xingx@microsoft.com}

\author{Meeyoung Cha}
\orcid{0000-0003-4085-9648}
\affiliation{%
  \institution{IBS \& KAIST}
  \city{Daejeon}
  \country{South Korea}
}

\email{mcha@ibs.re.kr}

\renewcommand{\shortauthors}{Sungwon Park et al.}

\begin{abstract}
Federated learning enables learning from decentralized data sources without compromising privacy, which makes it a crucial technique. 
However, it is vulnerable to model poisoning attacks, where malicious clients interfere with the training process.
Previous defense mechanisms have focused on the server-side by using careful model aggregation, but this may not be effective when the data is not identically distributed or when attackers can access the information of benign clients.
In this paper, we propose a new defense mechanism that focuses on the client-side, called \model{}, to help benign clients train robust local models and avoid the adverse impact of malicious model updates from attackers, even when a server-side defense cannot identify or remove adversaries. 
Our method consists of two main components: (1) attack-tolerant local meta update and (2) attack-tolerant global knowledge distillation.
These components are used to find noise-resilient model parameters while accurately extracting knowledge from a potentially corrupted global model. 
Our client-side defense strategy has a flexible structure and can work in conjunction with any existing server-side strategies.
Evaluations of real-world scenarios across multiple datasets show that the proposed method enhances the robustness of federated learning against model poisoning attacks.
\end{abstract}




\maketitle
\cutsectionup
\section{Introduction}

Federated learning has become a popular model training method to guarantee the minimum level of data privacy~\cite{bonawitz2019towards}. 
In each training round of federated learning, clients optimize their local models and send updates to the central server to aggregate them to produce a global model of the entire data distribution. Thus federated learning enables clients to jointly train a global model without directly sharing their private training data, making it a privacy-friendly solution~\cite{wang2021field}. 
Federated learning has been rapidly adopted by various applications that require data privacy~\cite{luo2021cost,rieke2020future,wu2022fedctr,park2022knowledge}.
\looseness=-1

Despite its advantages, federated learning is vulnerable to attacks due to its decentralized nature. Any client can easily participate in the training process, introducing the possibility of maliciousness~\cite{baruch2019little,lyu2020threats}. For example, federated learning systems assume that all participants are benign and that their data can help improve the performance of resulting models. This leaves room for \emph{model poisoning attacks}, wherein malicious users deceive the system by pretending to be benign and sending ``poisoned'' updates to the central server. 
This type of attack can disrupt parameter optimization~\cite{shafahi2018poison} and adversely impact the model's performance~\cite{gu2017badnets}, undermining the integrity of the federated learning system. 

Existing studies focus on using robust aggregation on the server side as a defense against model poisoning attacks.
A central server can be trained to preserve updates from likely benign clients while discarding updates from likely corrupted clients or adversaries.
Statistics like the trimmed mean or median can be used for outlier-resistant aggregation instead of simply averaging all updates.
Fu \textit{et al.,} extended this method by introducing a concept of confidence computed from residuals of repeated median estimator~\cite{fu2019attack}.
Other detection algorithms like Norm Bound, Multi-Krum, and FoolsGold assign lower weights to outlier updates and reduce their adverse effect~\cite{blanchard2017machine,fung2020limitations,sun2019can}.
However, these methods are limited in their ability to detect adversaries in the real world, where local datasets are no longer independent and identically distributed (i.e., non-IID). Non-IID makes benign local updates diverse and indistinguishable from corrupted ones~\cite{fung2020limitations}.
Furthermore, extant defenses have been breached by newer attacks that introduce elaborate local updates~\cite {baruch2019little,fang2020local,shejwalkar2021manipulating}, leading server-side strategies alone to be obsolete.

This paper takes a step further by introducing a client-side defense strategy named \model{}, which can run alongside existing server-side defense strategies to enhance resilience against poisoning attacks in federated learning.
This strategy modifies the local training process of benign clients by obtaining robust local models even when the server-side defense fails to filter out corrupt updates from malicious clients.

\model{} consists of two unique training components: (1) \textit{attack-tolerant local meta update} that finds local parameters that are less susceptible to noisy training by malicious clients and (2) \textit{attack-tolerant global knowledge distillation} that extracts the correct information from a noisy global model.

In our attack-tolerant local meta update component, \model{} generates a synthetic batch of corrupted data from the local data, including randomly flipped labels. 
The local model is then perturbed with this noisy batch of samples through one iteration update before the main objective is optimized using the clean batch of samples. 
This meta update is analogous to a vaccination process, where synthetic noise is introduced to prevent the model from collapsing in the face of attacks, and to find the model parameters that are more tolerant to noise. 
To generate more realistic noise during training, \model{} discovers $k$-nearest neighbors from the local model's embeddings to replace the original label. \looseness=-1

Meanwhile, the main objective of the attack-tolerant global knowledge distillation is to convey only useful knowledge from the possibly corrupted global model on the central server.
Conventional knowledge distillation cannot be used in federated learning attack scenarios because attackers' malicious influence on the corrupted global model can disrupt local training.
Instead, we leverage the fact that the last layer of a model is more prone to overfitting from noisy updates than the intermediate layer~\cite{kundu2021analyzing,stephenson2021geometry}. 
\model{} can learn more reliable information from the global model without overfitting by distilling knowledge only to the intermediate layer via an auxiliary network.
We also add a self-knowledge distillation objective to calibrate the prediction so that it works better for distillation.  \looseness=-1

Experiments show that \model{} is highly effective in various scenarios when applied in conjunction with existing server-side defense strategies.
Comparison with several recent baselines such as Norm bound, Multi-Krum, and ResidualBase shows that combining our method with these server-side strategies consistently results in better performance.
For instance, in the label poisoning attack scenario, the classification accuracy increased by 18\%, 20\%, 17\%, and 17\% for the CIFAR-10, CIFAR-100, TinyImageNet, and FEMNIST datasets, respectively.
We further show that \model{} is able to effectively defend against advanced poisoning attacks like LIE~\cite{baruch2019little}, STAT-OPT~\cite{fang2020local}, and DYN-OPT~\cite{shejwalkar2021manipulating}.

The major contributions and findings of our work include: 
\vspace{1mm}

\begin{itemize} [nosep,leftmargin=1em,labelwidth=*,align=left]

\item Proposing a unique client-side defense strategy, \model{}, that trains robust local models to thwart model poisoning attacks in federated learning. \looseness=-1

\item Designing an attack-tolerant local meta update that helps discover noise-tolerant parameters for local models by utilizing a synthetically corrupted training set.

\item Introducing an attack-tolerant global knowledge distillation technique that efficiently aligns the local model's knowledge to the global data distribution while reducing the adverse effects of false information in the possibly-corrupted global model.

\item Showcasing that \model{} can easily be applied in combination with any server-side defense strategies to enhance accuracy by 17-20\% under poisoning attacks across various datasets. 
\end{itemize}

\noindent 
The code and implementation details are available at the following URL: \url{https://github.com/deu30303/FedDefender/}.\looseness=-1

\cutsectionup
\section{Related Work}
\cutsubsectionup
\subsection{Poisoning Attacks in Federated Learning} 
In recent years, there is a growing concern about security issues in machine learning, leading to the emergence of various model poisoning attacks~\cite{chen2022federated,shejwalkar2022back}. 
Federated learning is particularly susceptible to these attacks because malicious users can easily access intermediate processes of the training and send poisoning updates to the central server~\cite{lyu2020threats, wu2022fedattack}.
Attackers can attack the training without access to the entire data distribution, as the data is decentralized and cannot be shared among clients~\cite{fang2020local,sikandar2023detailed}.

Model poisoning attacks can be divided into two categories: targeted attacks and untargeted attacks~\cite{gu2017badnets,shafahi2018poison}.
In targeted attacks, malicious clients inject a backdoor into the central server so that any instance with the backdoor trigger will be classified as ``targeted'' without degrading the overall performance of the model. 
In untargeted attacks, on the other hand, attackers aim to degrade model performance indiscriminately across all classes. 
A simple and widely used untargeted attack is the label-flipping attack~\cite{xiao2012adversarial}, in which malicious clients corrupt local models by randomly flipping labels of training samples from original classes to other classes. \looseness=-1

Recently, some studies have used benign clients' partial information to improve the stealthiness of untargeted attack performance. 
For example, Baruch \textit{et al.} infer the standard deviation and the intensity factor based on benign client's updates and inject perturbed model updates accordingly~\cite{baruch2019little}.
Fang \textit{et al.} inject malicious updates by adding constant opposite direction noise estimated from benign clients' updates mean~\cite{fang2020local}. 
Shejwalkar \textit{et al.} propose a model poisoning attack by calculating a dynamic malicious update opposite to the direction of benign model updates~\cite{shejwalkar2021manipulating}. 

This paper focuses on defending against \textit{untargeted attacks} in federated learning using the client-side defense.  \looseness=-1
\begin{figure*}[t!]
\centerline{
   \includegraphics[width=0.85\linewidth]{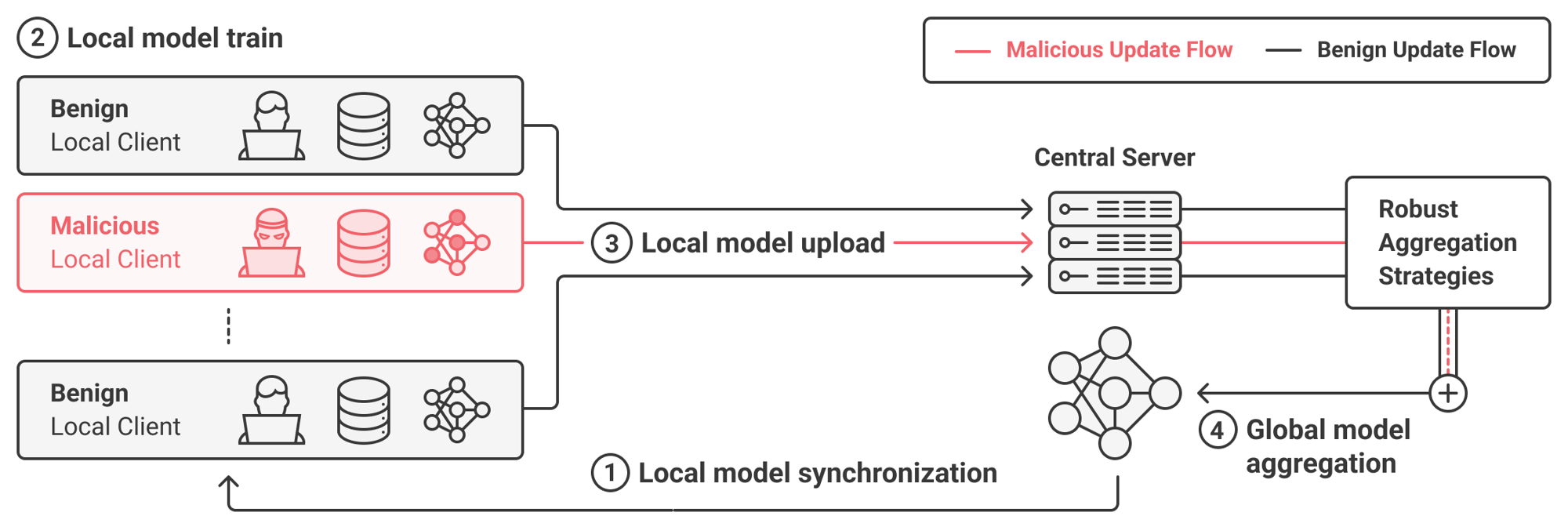}}
   \caption{Illustration of federated learning under poisoning attacks. Attackers attempt to send malicious updates to the central server to corrupt the aggregated model. Unlike existing works that primarily focus on robust aggregation on the server side (stage \textcircled{\footnotesize{4}}), \model{} focuses on training robust local models by benign clients (stage \textcircled{\footnotesize{2}}) to protect against malicious updates from adversaries. \looseness=-1
   } 
\label{fig:fedavg}
\end{figure*}
\cutsubsectionup
\subsection{Server-Side Defenses Against Poisoning Attacks in Federated Learning} 
In federated learning, dimension-wise averaging is a commonly used and effective method for aggregating local updates~\cite{chen2016revisiting,mcmahan2017communication}. 
However, this naive averaging method is vulnerable to model poisoning attacks, which may succeed even with just a single malicious model update~\cite{awan2021contra,blanchard2017machine}.
To address this threat, two server-side defense strategies have been proposed: coordinate-wise aggregation and detection-wise aggregation. Both aim to filter out malicious updates from adversary clients during the aggregation and ensure the integrity of the federated learning process.

\cutparagraphup
\noindent\textbf{Coordinate-wise Aggregation.} 
Coordinate-wise aggregation introduces outlier-resistant operations instead of averaging. 
For example, \textit{Trimmed Mean} eliminates the largest and smallest values in each dimension, before computing the mean~\cite{yin2018byzantine}.
\textit{Median} is another dimension-wise aggregation algorithm that computes the median of the updates in place of averaging~\cite{xie2018generalized}. 
However, both approaches are ineffective when the data distribution is non-IID as they may overlook underrepresented updates. 
To tackle this limitation, \textit{ResidualBase} proposes a residual-based aggregation method~\cite{fu2019attack}, which calculates residuals of each parameter in the local model using a repeated median estimator. 
These residuals are then used to determine parameter confidence and detect false updates.
 
\cutparagraphup
\noindent\textbf{Detection-wise Aggregation.} 
To mitigate the adverse effect of malicious clients' updates, detection-wise aggregation adjusts learning rates based on the abnormality score of each local update.
\textit{Norm Bound} disregards clients with the norm of local updates being above a certain threshold by exploiting the observation that malicious clients often produce updates with a larger variance and norm than benign clients~\cite{sun2019can}. 
\textit{Krum} introduces a Byzantine-resilient algorithm assuming that malicious updates would be placed far from benign updates in the Euclidean space~\cite{blanchard2017machine}.
More specifically, it selects a single local client update that is the most similar to its $n-m-2$ neighboring updates to produce the global model, where $m$ is the expected number of malicious local clients. 
Krum can be extended to Multi-Krum by iteratively running the algorithm to select multiple local updates, which shows better robustness than the original Krum. 
\textit{FoolsGold} identifies grouped actions of targeted attacks based on the similarity score among local updates~\cite{fung2020limitations}. 
Unlike benign clients, malicious clients in a targeted attack scenario share the common loss objective and tend to have a more similar update pattern than benign clients. 
In this context, FoolsGold adjusts the learning rates of local models in proportion to the degree of diversity in each local update.

While the client-side defense has been relatively under-investigated, FL-WBC proposed a client-side defense strategy specifically designed to mitigate backdoor attacks~\cite{sun2021fl}. In contrast, our proposed method, \model{}, stands as a pioneering client-side defense strategy against untargeted attacks in federated learning. By complementing existing server-side defense strategies, it significantly enhances the overall resilience against model poisoning attacks when used in conjunction with them. \looseness=-1
\label{sec:method}
\cutsectionup
\section{Problem Formulation}
 \begin{figure*}[t!]
\centerline{
      \includegraphics[width=1\linewidth]{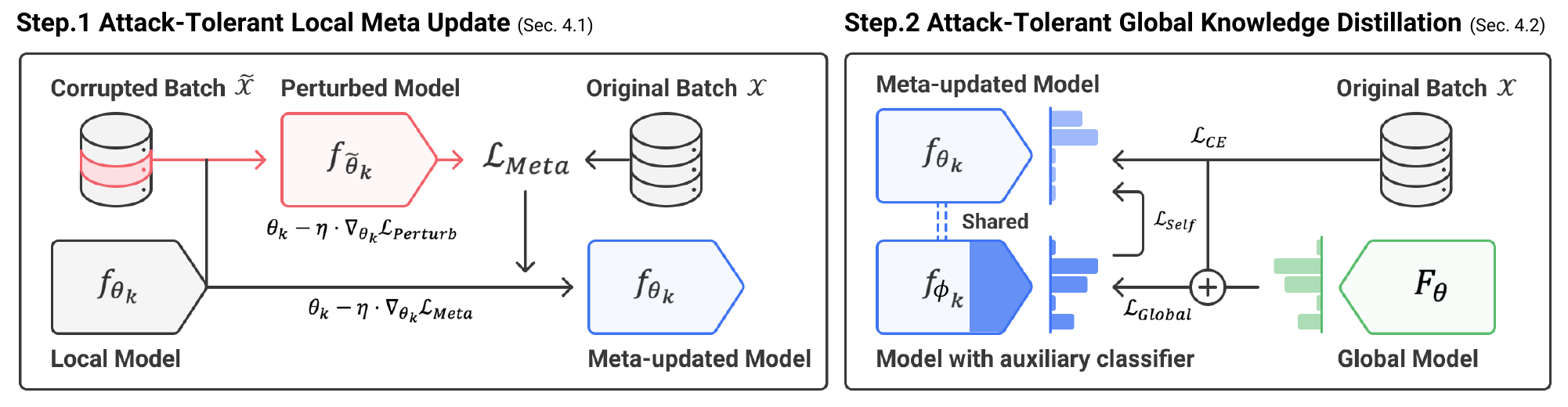}}
      \caption{The overall design of \model{} framework. \model{} comprises two steps: (1) attack-tolerant local meta-update, which finds local model parameters that are less prone to overfitting to noise, and (2) attack-tolerant global knowledge distillation, which aims to convey correct global knowledge from a potentially contaminated global model, regularizing the local model to mitigate data bias. \looseness=-1
      } 
\label{fig:main_model}
\end{figure*}
We will outline model poisoning attacks in federated learning and defenses against them in this section. 
Figure~\ref{fig:fedavg} highlights the basic flow of the training process in federated learning when malicious attackers are present. \looseness=-1
\cutsubsectionup
\subsection{Federated Learning}
\cutparagraphup \noindent
\textbf{Overview.~}  Federated learning aims to optimize a global model, denoted as $F_{\theta}$, by utilizing data distributed across $N$ local clients. 
We denote the loss objective and local dataset in the $k$-th client by $\mathcal{L}_{k}$ and $\mathcal{D}_{k}$, respectively. The primary objective for training $\theta$ can be expressed as: \looseness=-1
\begin{align}
    \min_{\theta} \mathcal{L}(\theta) = \min_{\theta} {\sum_{k=1}^{N} |\mathcal{D}_k| \cdot \mathcal{L}_k(\theta, \mathcal{D}_{k}) \over \sum_{k=1}^{N} |\mathcal{D}_k|}.
    \label{eq:fed_objective}
\end{align}

In this paper, we use FedAvg~\cite{mcmahan2017communication}, a widely used aggregation method as the default setting. FedAvg comprises four stages in each communication round:
\begin{enumerate}[leftmargin=*,label=\large\protect\textcircled{\footnotesize\arabic*}]
\item Local model synchronization: At the beginning of each round $t$, a random subset of local clients is selected, and these clients synchronize their model parameters by downloading the current global model parameter $\theta^t$.

\item Local model train: Each selected client $k$ trains the downloaded model with her private data $\mathcal{D}_{k}$ for a few epochs and 
computes the local model update $\Delta \theta_k^t = \theta_k^{t} - \theta^{t}$.

\item Local model upload: The local model update $\Delta \theta_k^t$ is sent to the central server.

\item Global model aggregation: The global model is updated by averaging all local updates received from the clients at the central server, as shown in \eqref{eq:global_aggregation}:
\end{enumerate}
\begin{align}
    \theta^{t+1} = \theta^t + \sum_{k=1}^{N} {|\mathcal{D}_k| \over |\mathcal{D}|} \Delta \theta_k^t, 
    \label{eq:global_aggregation}
\end{align}
where $|\mathcal{D}|=\sum_{k=1}^N |\mathcal{D}_k|$. This process is repeated for each communication round. \looseness=-1

\cutparagraphup \noindent
\textbf{Threat model.~} We hypothesize that all benign clients are honest and follow the proposed protocol, whereas malicious clients do not. Adversaries aim to disrupt the convergence and deteriorate the performance of the global model (i.e., an untargeted attack) by poisoning the local models of malicious clients. We also assume that the central server is trustworthy and performs the required task faithfully.
Regarding the attacker's capability, we assume that adversaries can bypass the verification process and infiltrate the federated system as participating clients (e.g., during optimization). We consider two scenarios for the threat model: (1) attackers have no access to benign clients' information, and (2) attackers have partial access to benign client information, such as the standard deviation or average updates from benign clients.

\cutsubsectionup
\subsection{Federated Learning When Facing Poisoning Attacks}

Poisoning attacks rely on sending corrupted local updates to the central server, ultimately contaminating the trained global model.
Several malicious objectives can be used for such attacks.
For instance, attackers may adopt the same loss objective as legitimate clients for optimization, but use deliberately corrupted training set $\Tilde{\mathcal{D}}_k$ to convey false information (i.e., $\mathcal{L}_k (\theta, \Tilde{\mathcal{D}}_k)$).
Additionally, attackers can also introduce an adversarial objective $\Tilde{\mathcal{L}}_k$ that disrupts the overall convergence (i.e., $\Tilde{\mathcal{L}}_k (\theta, \mathcal{D}_k)$). \looseness=-1

We denote the subset of malicious clients as $\mathcal{S}_m$ and the subset of benign clients as $\mathcal{S}_b$ from the entire set of $N$ clients $\mathcal{S}$ (i.e., $\mathcal{S}$ = $\mathcal{S}_m$ $\cup$ $\mathcal{S}_b$).
To defend against poisoning attacks, the optimal objective for training the global model weight $\theta$ can be defined as:
\begin{align}
   \min_{\theta} \mathcal{L}^*(\theta) &=  \min_{\theta} { \sum_{k \in \mathcal{S}_b} |\mathcal{D}_k| \cdot \mathcal{L}_k(\theta, \mathcal{D}_{k}) \over \sum_{k \in \mathcal{S}_b} |\mathcal{D}_k|}.
    \label{eq:fed_optimal_objective}
\end{align}

One approach to achieving this objective is to implement a server-side robust aggregation strategy for the global model aggregation (i.e., stage \textcircled{\footnotesize{4}}), which aims to retain the updates from benign clients while discarding all false updates from malicious clients during the global aggregation phase.
The objective of robust aggregation can be expressed as follows: \looseness=-1
\begin{align}
\theta^{t+1} &= \theta^t + {{\sum_{k=1}^N \mathbbm{1}_{\{k \in S_{b}\}}} \cdot \Delta \theta_{k}^{t} \over |S_b|},
\label{Eq:optimal_aggregation}
\end{align}
\noindent where $\mathbbm{1}$ is an indicator function. $|\mathcal{D}_k|$ is removed in order to mitigate the attack scenario where attackers try to manipulate the size of the local training data.

In contrast to existing defenses that focus on robust aggregation strategies, \model{} aims to directly solve \eqref{eq:fed_optimal_objective} by redesigning the local model train process (stage \textcircled{\footnotesize{2}}) and modifying the local updates of legitimate clients $\Delta \theta_{k}^{t}$. The detail is described in the next section. \looseness=-1

\section{Description of Our Method}
%\textbf{Overview. } 
\model{} comprises the following two steps,% attack-tolerant training update for local models at each mini-batch $\mathcal{X}$.
\begin{itemize}[leftmargin=*]
    \item \noindent \textbf{Step 1. Attack-Tolerant Local Meta Update (Section~\ref{Sec:local_update})}:   
    We train the benign local model $f_{\theta_k}$ in a robust manner via meta-learning. The goal is to discover the model's parameters that produce accurate predictions even after it has been perturbed by noisy information.
    To achieve this, we first generate a noisy synthetic label, then apply one gradient update to perturb the local network parameters. Afterward, the gradient for the perturbed network to predict the correct outputs is computed and utilized to optimize the original local model. 

    \cutparagraphup
    \item \noindent \textbf{Step 2. Attack-Tolerant Global Knowledge Distillation (Section~\ref{Sec:global_update})}: 
    If the global aggregation defense cannot block updates from malicious clients, the global model $F_{\theta}$ is no longer trustworthy. 
    Given the meta-updated local model from the previous step, we apply global knowledge distillation to an auxiliary classifier using intermediate feature maps to neutralize the adverse impact on the contaminated global model $F_\theta$. 
    Self-knowledge distillation is applied between the auxiliary classifier and the original classifier to further incorporate the global knowledge into the deeper layers of the local model. \looseness=-1
    %to improve the performance of the local model and incorporate the calibrated global knowledge. \looseness=-1

\end{itemize}

Figure~\ref{fig:main_model} present the overall pipeline of our proposed defense. Our technique can mitigate model poisoning attacks in federated learning. We will now provide a detailed description of each component of \model{} next.

\cutsubsectionup
\subsection{Attack-Tolerant Local Meta Update} \label{Sec:local_update}

If the local model parameters $\theta_{k}$ are only optimized with a traditional supervised loss term (e.g., cross-entropy), it is susceptible to overfitting and being influenced by noise generated by malicious clients. 
Our key idea is to learn noise-tolerant parameters $\theta_{k}$ in a way that "vaccinates" the local model $f_{\theta_{k}}$ against model poisoning with synthetic noise, drawing inspiration from recent meta-learning works~\cite{li2019learning, finn2017model, ravi2017optimization}.
The proposed local meta-update replicates the training context with a model poisoning attack and makes the network less sensitive to noisy perturbations. \looseness=-1

\cutparagraphup \noindent\textbf{Local model poisoning with synthetic noise.} 
Let us denote a mini-batch from local dataset $\mathcal{D}_{k}$ as $\mathcal{X} = \{(\mathbf{x}_i, \mathbf{y}_i)\}_{i=1}^B$, where $\mathbf{x}$ is an input instance and $\mathbf{y}$ is the corresponding one-hot label. 
We want to generate synthetic batches $\mathcal{\Tilde{X}} = \{(\mathbf{x}_i, \mathbf{\Tilde{y}}_i)\}_{i=1}^B$ with label noise to simulate the poisoning attack and perturb the local model.

Excessive perturbation can overly deform the model's decision boundary, leading to degraded performance. 
To mitigate this risk of severe deformation, we create more realistic noisy labels that resemble the distribution of $\mathbf{y}$ by transferring labels from similar samples.
For each $(\mathbf{x},\mathbf{y})\in\mathcal{X}$, we calculate its feature representation $h_\mathbf{x}$ from the model's backbone network.
Then, a random instance from top-$k$ nearest neighbors in the representation space is randomly selected to replace the original label: \looseness=-1
\begin{align}
\mathcal{\Tilde{X}} &= \{(\mathbf{x}, \mathbf{\Tilde{y}}) | (\mathbf{x}, \mathbf{y}) \in \mathcal{X} \text{ and }\mathbf{\Tilde{y}} = \text{Sample}_{\mathbf{y}}(\mathcal{N}_{k}(\mathbf{x}, \theta_k))  \},
%\mathcal{\Tilde{X}} &= \text{Perturb}(\mathcal{X})
\end{align}
where $\mathcal{N}_{k}(\mathbf{x}, \theta)$ indicates the set of $k$-nearest neighbors of $\mathbf{x}$ from the representation space made by $f_\theta$. 
$\text{Sample}_{\mathbf{y}}(\cdot)$ is the random selector function to extract the synthetic label.
Since the nearest neighbors $\mathcal{N}_{k}(\mathbf{x}, \theta)$ are likely to share the same label as $\mathbf{x}$, this can generate noise within an acceptable range (i.e., 5-20\% error rate).

Given the synthetic batch samples $\mathcal{\Tilde{X}}$, we perturb the local model parameter $\theta_{k}$ using one gradient descent step:
\begin{align}
\mathcal{L}_{Perturb} &= {1 \over |\mathcal{\Tilde{X}}|} \sum_{{\mathbf{x}}, {\mathbf{\Tilde{y}}} \in \mathcal{\Tilde{X}}} H(\mathbf{\Tilde{y}}, f_{\theta_k}({\mathbf{x}})) \\
\Tilde{\theta}_k &\leftarrow \theta_k - \eta \nabla_{\theta_k}  \mathcal{L}_{Perturb},
\end{align}
where $H(p, q)$ is the conventional cross entropy between $p$ and $q$, and $\eta$ is a learning rate. \looseness=-1

\cutparagraphup \noindent\textbf{Local model correction with meta update.}
To discover the model parameter that is less susceptible to noisy training by malicious clients, we propose a meta update to guide the model training.
Given the perturbed model $\Tilde{\theta}_k$, we optimize a classification loss $\mathcal{L}_{Meta}$ (\eqref{eq:meta_loss}) for one gradient descent step to encourage the perturbed local model $f_{\Tilde{\theta}_k}$ to give
correct predictions from $\mathcal{X}$.
Note that the optimization process is applied to the parameters of the original local model $\theta_{k}$ (\eqref{eq:meta_update}), although the loss calculation is based on the perturbed model's parameters $\Tilde{\theta}_k$ (\eqref{eq:meta_loss}).
This prevents the local model from being contaminated by synthetic noise during the training.
\begin{align}
\mathcal{L}_{Meta} &= {1 \over |\mathcal{X}|} \sum_{{\mathbf{x}}, {\mathbf{y}} \in \mathcal{X}} H(\mathbf{y}, f_{\Tilde{\theta}_k}({\mathbf{x}})) \label{eq:meta_loss} \\
\theta_k &\leftarrow \theta_k - \eta \nabla_{\theta_k}  \mathcal{L}_{Meta} \label{eq:meta_update}
\end{align} \looseness=-1

\subsection{Attack-Tolerant Global 
Knowledge Distillation}\label{Sec:global_update}
Given the meta-updated network $f_{\theta_{k}}$, our next step is to enhance the performance of the refined local model $\theta_k$ through global knowledge distillation. 
In the case of non-IID settings, learning relying on local data leads to representation bias, as local data distribution differs from the global distribution. 
Regularization techniques, including global knowledge distillation, can be employed to control the local updates, thus preventing the local drift fallacy~\cite{han2022fedx, zhuang2020performance,wu2022communication}.

In the scenario of a model poisoning attack, however, the credibility of the global model $F_{\theta}$ can be compromised if the global defense fails to block malicious updates from hostile clients. 
This can lead to suboptimal results when the $F_{\theta}$ is perturbed.
We propose attack-tolerant global knowledge distillation to mitigate the adverse effects of a potentially corrupted global model $F_{\theta}$. \looseness=-1

\cutparagraphup \noindent\textbf{Refined knowledge distillation with an auxiliary network.} 
We start with the fact that a deeper layer in deep neural networks is much easier to overfit to noise (i.e., memorization) due to the inherent nature of gradient descent-based optimization~\cite{kundu2021analyzing, baldock2021deep, stephenson2021geometry}.
In this context, \model{} transfers global knowledge to a shallow intermediate part of the local model to reduce the adverse effect of false information.
We attach an auxiliary classifier on top of intermediate layers to produce a new model parameter $\phi_k$, and perform knowledge distillation on it (i.e., $\theta_{k}$ and $\phi_k$ share the shallow section of the entire network). Given an input data point $\mathbf{x}$ from mini-batch $\mathcal{X}$, we regard the output probability from the global model $F_\theta$ as a target to train the local model parameter with auxiliary classifier $\phi_k$. The global knowledge distillation loss is defined as follows:
 \begin{align}
 &F_\theta (\mathbf{x}, \tau) = \text{Sharpen}(F_\theta(\mathbf{x}), \tau) \\
    &\mathcal{L}_{Global} = {1 \over |\mathcal{X}|} \sum_{{\mathbf{x}}, {\mathbf{y}} \in \mathcal{X}} H(F_{\theta}(\mathbf{x}, \tau), f_{\phi_{k}}({\mathbf{x}})),
\end{align} 
where Sharpen($\cdot$) is the function that adjusts the confidence with sharpening temperature $\tau$~\cite{hinton2015distilling}.

Since the global model's prediction is still not reliable in the attack scenario, we further introduce a simple strategy to refine target probabilities for robust knowledge distillation.
Given a data point ($\mathbf{x}, \mathbf{y}$) from $\mathcal{X}$, we calculate the scale coefficient $\alpha$, which represents the cosine similarity between original label $\mathbf{y}$ and the global model prediction $F_{\theta}(\mathbf{x})$.
\model{} then generates a refined label $\mathbf{\hat{y}}$ by applying a linear sum to calibrate the global information as follows:
\begin{align}
\mathbf{\hat{y}} = (1 - \alpha) \cdot \mathbf{y} + \alpha\cdot F_{\theta}(\mathbf{x}, \tau). \label{eq:refine_label}
\end{align}
If the global model's prediction is well aligned with the ground-truth labels, the scale coefficient becomes large and the final label weighs more on the global model's knowledge.
Meanwhile, if the prediction is far different from the truth, we neglect the global model's information.
We construct a modified batch $\mathcal{\hat{X}}$ with the refined labels $\mathbf{\hat{y}}$, and transfer the global knowledge to $\phi_k$ with this batch (\eqref{eq:global_kd_real}).
\begin{align}
    \mathcal{L}_{Global} = {1 \over |\mathcal{\hat{X}}|} \sum_{{\mathbf{x}}, {\mathbf{\hat{y}}} \in \mathcal{\hat{X}}} H(\mathbf{\hat{y}}, f_{\phi_{k}}({\mathbf{x}})) \label{eq:global_kd_real}
\end{align} 
\looseness=-1

\cutparagraphup \noindent\textbf{Auxiliary self-knowledge distillation.} 
We propose an additional design to improve the deeper layers of the local model, whose weights are excluded from global knowledge distillation, in learning calibrated global knowledge.
Motivated by the notion of self-knowledge distillation~\cite{zhang2019your, park2021improving,han2020mitigating}, we extract knowledge from the local model using an auxiliary classifier $\phi_k$ and transfer it back to the original local model $\theta_k$. 
By doing so, we can distill the calibrated global knowledge learned from the refined label (\eqref{eq:refine_label}) to the latter part of the model, effectively regularizing the entire model.
Moreover, the output from the auxiliary head can be considered as a different view (i.e., augmentation) of the same instance.
Maximizing the agreement between these two views can further enhance the model training.
The self-knowledge distillation loss between the auxiliary classifier $f_{\phi_k}$ and the original model $f_{\theta_k}$ is defined as follows.
\begin{align}
\mathcal{L}_{Self} &= {1 \over |\mathcal{X}|} \sum_{{\mathbf{x}}, {\mathbf{y}} \in \mathcal{X}} KL( f_{\theta_k}({\mathbf{x}, \tau}) || f_{\phi_k}({\mathbf{x}, \tau})), 
\end{align}
\noindent where $KL(p||q)$ is a Kullback–Leibler (KL) divergence between two probabilities $p$ and $q$.
Finally, the final loss for our global knowledge distillation is given in \eqref{eq:our_global_kd}.
\begin{align}
 \mathcal{L}_{KD} &= \mathcal{L}_{Global} + \mathcal{L}_{Self} \label{eq:our_global_kd}
\end{align}

This global knowledge distillation loss is optimized in conjunction with original cross-entropy loss to train the \model{} (\eqref{eq:total_training}).
\begin{align}
&\mathcal{L}_{CE} = {1 \over |\mathcal{X}|} \sum_{{\mathbf{x}}, {\mathbf{y}} \in \mathcal{X}} H(\mathbf{y}, f_{\theta_k}({\mathbf{x}})) \\
&\mathcal{L}_{total} = \mathcal{L}_{CE} + \mathcal{L}_{KD} \\
&\theta_k \leftarrow \theta_k - \eta \nabla_{\theta_k}  \mathcal{L}_{total} \label{eq:total_training}
\end{align}
\looseness=-1

\cutsectionup
\section{Experiments}
We evaluate the robustness of \model{} under various attacks on multiple datasets and compare it with server-side global aggregation baselines. We also examine how the model components and hyperparameters affect the overall performance. \looseness=-1

\cutsubsectionup
\subsection{Experimental Setup}
\cutparagraphup
\noindent \textbf{Data settings.} We use four image classification benchmark datasets in our experiments. The first two are CIFAR-10 and CIFAR-100, each containing 60,000 images of 32x32 pixels~\cite{krizhevsky2009learning}. 
CIFAR-10 comprises 10 classes, such as airplanes, cats, and dogs, while CIFAR-100 comprises 100 classes. The third dataset is TinyImageNet, which comprises 100,000 images of 64x64 pixels and 200 classes. 
The fourth is the Federated Extended MNIST (FEMNIST) dataset, which contains 805,263 images of handwriting digits/characters of the size of 28x28 pixels~\cite{caldas2018leaf}. 

\begin{table}[t]
{
\setlength{\tabcolsep}{2.5pt}
\renewcommand{\arraystretch}{1.2}
\caption{Performance improvement with \model{} on classification accuracy in Scenario-1 over four datasets. Our model brings non-trivial improvement for all server-side baseline algorithms for both last and best accuracy.}
\cutcationup
\label{tab:main_result}
\resizebox{1.0\linewidth}{!}{
\begin{tabular}{l|cc|cc|cc|cc}
\toprule
\multirow{2}{*}{Method} & \multicolumn{2}{c|}{CIFAR-10} &  \multicolumn{2}{c|}{CIFAR-100} & \multicolumn{2}{c|}{TinyImageNet} & \multicolumn{2}{c}{FEMNIST} \\ 
                      & ~~ Last ~~  & ~~ Best ~~  &  ~~ Last ~~ & ~~ Best ~~  & ~~ Last ~~ & ~~ Best ~~ & ~~ Last ~~ & ~~ Best ~~  \\ \midrule
No Defense            &   68.80     &  71.96   &  42.97   & 43.90  & 30.37  & 38.98 & 18.88 & 23.81  \\
+ \model{}           & \textbf{78.17} & \textbf{79.96} & \textbf{51.76}  & \textbf{51.92} &  \textbf{35.59}  &  \textbf{39.68}  & \textbf{22.11} & \textbf{24.48} \\ \hline
Median               &	62.02 & 64.76 & 36.33	& 37.54 & 26.73 & 32.53 &12.00 & 20.03 \\
+ \model{}            	&  \textbf{72.96} & \textbf{76.28} &  \textbf{39.89} & \textbf{41.29}  & \textbf{26.92} & \textbf{32.88} & \textbf{15.75}  & \textbf{21.24}  \\ \hline
Trimmed Mean                &  75.52      & 76.03     &  47.93  & 47.95  &  36.76 & 38.24 & 19.37 & 23.17\\
+ \model{}           	  &    \textbf{81.06}    &   \textbf{81.96}   &  \textbf{52.90}  & \textbf{53.41}  & \textbf{38.85}  &   \textbf{39.37} & \textbf{21.80} & \textbf{23.95}\\ \hline
Norm Bound          	  &  67.36       &  70.07    &  42.51  & 45.12  &  30.78 & 35.87 & 18.50 & 23.22 \\
+ \model{}             &  \textbf{74.05}  &   \textbf{75.84}     &   \textbf{48.39}   &  \textbf{49.09}  & \textbf{31.69}  & \textbf{36.17}  & \textbf{20.55} & \textbf{24.36} \\ \hline
Multi-Krum                   &  73.09      &   75.03   &  47.75  & 47.83  &  37.26 &  38.54 & 20.55 & 23.30       \\
+ \model{}                & \textbf{81.87}  &  \textbf{82.77}     & \textbf{53.15}     &  \textbf{53.35}  &    \textbf{38.98}   & \textbf{39.48} & \textbf{22.43} & \textbf{24.36}\\ \hline
ResidualBase               & 73.61  & 75.10      &  44.80    & 45.13  &   35.05     &  38.60 & 19.44 & 23.86\\
+ \model{}             &  \textbf{79.28} & \textbf{80.83}      &  \textbf{50.62}    &  \textbf{50.98}  &  \textbf{36.22}     &  \textbf{39.24} & \textbf{22.41} & \textbf{24.27}  \\ \bottomrule
\end{tabular}}
}
\end{table}
\begin{table*}[t!]
\caption{Performance improvement with \model{} on classification accuracy in Scenario-2 over four datasets. Our model brings non-trivial improvement for all server-side baseline algorithms for both last and best accuracy.}
\label{tab:main_result2}
\cutcationup
{
\begin{tabular}{l|cc|cc|cc}
\toprule
\multirow{2}{*}{CIFAR-10} & \multicolumn{2}{c|}{LIE} &  \multicolumn{2}{c|}{STAT-OPT} & \multicolumn{2}{c}{DYN-OPT}  \\ 
                      &  Last &  Best &  Last &  Best  & Last & Best   \\ \midrule
No Defense            &   73.40  & 76.68  & 73.74   & 73.99 &  62.37 & 63.13\\
+ \model{}           & \textbf{79.69} & \textbf{82.78} & \textbf{79.77}  & \textbf{80.46} &  \textbf{68.78}  &  \textbf{69.53} \\ \hline
Median               & 55.71 &  64.80 & 53.37	& 58.17 & 74.71 & 76.65 \\
+ \model{}            	& \textbf{60.58}  & \textbf{72.97} & \textbf{60.88}  & \textbf{66.01} & \textbf{81.86} & \textbf{83.07} \\ \hline
Trimmed Mean                &   39.19  &  76.29  & 64.46 &  68.34 & 72.80  & 75.18 \\
+ \model{}           	  &  \textbf{42.86}   &  \textbf{79.37}   &  \textbf{82.53}  & \textbf{83.05}  & \textbf{81.29}  &  \textbf{82.18} \\ \hline
Norm Bound          	  &   12.30    & 27.90 & 64.15  & 64.70 & 72.54  & 75.09\\
+ \model{}             & \textbf{14.00}  & \textbf{38.34}     &  \textbf{82.45}  & \textbf{82.85} & \textbf{81.47} & \textbf{82.42}\\ \hline
Multi-Krum                   &   41.46    & 75.80   & 75.56    & 76.10  & 71.49  &  75.25    \\
+ \model{}                & \textbf{46.30} & \textbf{79.05}  &  \textbf{82.33} & \textbf{82.80}   & \textbf{80.94}   & \textbf{82.43} \\ \hline
ResidualBase               &  74.53 &  77.04    &  70.24  & 70.39 &     74.40  & 76.93 \\
+ \model{}             & \textbf{83.19}  &  \textbf{83.65}     &   \textbf{77.81}   & \textbf{78.73}  &   \textbf{83.48} &  \textbf{83.90} \\ \bottomrule
\end{tabular}
}
\begin{tabular}{l|cc|cc|cc}
\toprule
\multirow{2}{*}{CIFAR-100} & \multicolumn{2}{c|}{LIE} &  \multicolumn{2}{c|}{STAT-OPT} & \multicolumn{2}{c}{DYN-OPT}  \\ 
                       &  Last &  Best &  Last &  Best  & Last & Best   \\ \midrule
No Defense            &  48.99  & 49.43  & 23.40   & 24.16 &  49.33 & 49.70\\
+ \model{}           & \textbf{52.88} & \textbf{53.46} & \textbf{29.49}  & \textbf{29.49} &  \textbf{53.86}  & \textbf{54.21}  \\ \hline
Median               & 46.40 &  47.12 & 16.95	& 17.02 & 46.81 & 46.91 \\
+ \model{}            	& \textbf{50.46}  & \textbf{52.25} & \textbf{23.21}  & \textbf{23.62} & \textbf{51.11} & \textbf{51.49} \\ \hline
Trimmed Mean                &   47.45 &  47.77  & 25.24 &  25.27 & 47.80 & 47.87 \\
+ \model{}           	  &  \textbf{51.48}   &  \textbf{51.91}   &  \textbf{29.80}  & \textbf{29.89}  & \textbf{52.64}  & \textbf{52.86} \\ \hline
Norm Bound          	  &   47.42    & 47.77 & 23.61  & 23.66 & 47.83 & 48.01\\
+ \model{}             & \textbf{52.15}  & \textbf{52.76}     &  \textbf{30.45}  & \textbf{30.60} & \textbf{53.09} & \textbf{53.19}\\ \hline
Multi-Krum                   &   48.23    & 48.66   & 25.34   & 25.62  &  48.23 & 48.41     \\
+ \model{}                & \textbf{51.79} & \textbf{52.57}  &  \textbf{31.42} & \textbf{31.52}   & \textbf{52.96}   & \textbf{53.23} \\ \hline
ResidualBase               &  48.68 &  48.88    &  26.46  & 26.57 &     48.96  & 49.03 \\
+ \model{}             & \textbf{53.11} &  \textbf{53.55}     &   \textbf{31.41}   & \textbf{31.41}  &   \textbf{54.39} &  \textbf{54.70} \\ \bottomrule
\end{tabular}
\begin{tabular}{l|cc|cc|cc}
\toprule
\multirow{2}{*}{TinyImageNet} & \multicolumn{2}{c|}{LIE} &  \multicolumn{2}{c|}{STAT-OPT} & \multicolumn{2}{c}{DYN-OPT}  \\ 
                      &  Last &  Best &  Last &  Best  & Last & Best   \\ \midrule
No Defense            &   35.07  & 37.92  & 26.82   & 26.82 &  30.66 & 30.66\\
+ \model{}           & \textbf{37.61} & \textbf{39.27} & \textbf{29.62} &  \textbf{29.62}  & \textbf{32.50}  &  \textbf{32.52} \\ \hline
Median               & 26.79 &  36.67 & 18.27	& 18.39 & 20.12 & 20.33 \\
+ \model{}            	& \textbf{33.42}  & \textbf{36.90} & \textbf{21.68}   &  \textbf{23.63} & \textbf{23.65} & \textbf{23.89} \\ \hline
Trimmed Mean                &   33.15  &  35.87  & 26.52 & 26.54 & 26.40  & 26.47 \\
+ \model{}           	  &  \textbf{35.70}   &  \textbf{37.28}   &  \textbf{27.26}  & \textbf{27.29}  & \textbf{31.42}  & \textbf{31.48} \\ \hline
Norm Bound          	  &   32.43   & 35.45 & 23.61 & 23.70 & 27.17 & 27.19\\
+ \model{}             & \textbf{35.59}  & \textbf{37.42}     &  \textbf{26.66}  & \textbf{26.66} & \textbf{31.45} & \textbf{31.49}\\ \hline
Multi-Krum                   &   33.38    & 36.32   & 26.39  & 26.40  &  26.31 & 26.31      \\
+ \model{}                & \textbf{35.22} & \textbf{37.04}  & \textbf{27.83}  & \textbf{27.95} & \textbf{31.34}   & \textbf{31.44} \\ \hline
ResidualBase               &  33.61 &  38.21    &  28.10  & 28.10 &     27.93  & 
 27.96 \\
+ \model{}             & \textbf{36.73}  &  \textbf{38.73}     &  \textbf{28.49}   & \textbf{28.51}  &   \textbf{31.85} &  \textbf{31.88} \\ \bottomrule
\end{tabular}
\begin{tabular}{l|cc|cc|cc}
\toprule
\multirow{2}{*}{FEMNIST} & \multicolumn{2}{c|}{LIE} &  \multicolumn{2}{c|}{STAT-OPT} & \multicolumn{2}{c}{DYN-OPT}  \\ 
                      &  Last &  Best &  Last &  Best  & Last & Best   \\ \midrule
No Defense            &   20.07  & 22.30  & 19.67   & 22.43 &  19.41 & 22.63\\
+ \model{}           & \textbf{21.87} & \textbf{24.03} & \textbf{21.68}  & \textbf{24.13} &  \textbf{21.42}  &  \textbf{24.29} \\ \hline
Median               & 17.81 &  23.15 & 10.56 & 14.60 & 19.14 & 22.52 \\
+ \model{}            	& \textbf{20.58}  & \textbf{23.39} & \textbf{15.02}  & \textbf{19.16} & \textbf{21.01} & \textbf{24.16} \\ \hline
Trimmed Mean                &   19.01 &  22.21  & 19.48  & 22.83  & 18.19  & 21.71 \\
+ \model{}           	  &  \textbf{21.67}   &  \textbf{23.73}   &  \textbf{21.99} & \textbf{24.24}  & \textbf{20.70} &  \textbf{23.77} \\ \hline
Norm Bound          	  &   20.50    & 22.70 & 19.14 & 21.97 & 19.55 & 22.25\\
+ \model{}             & \textbf{21.64}  & \textbf{23.79}     &  \textbf{21.59}  & \textbf{24.15} & \textbf{20.89} & \textbf{23.88}\\ \hline
Multi-Krum                   &   20.25   & 22.43   & 19.88    & 23.16  &  19.56 & 22.82     \\
+ \model{}                & \textbf{21.93} & \textbf{23.44}  &  \textbf{22.06} & \textbf{24.16}   & \textbf{20.96}   & \textbf{23.87} \\ \hline
ResidualBase               &  20.21 &  22.92    &  19.07 & 21.53 &    19.28  & 23.08 \\
+ \model{}             & \textbf{22.66}  &  \textbf{24.01}     &   \textbf{22.01}   & \textbf{23.97}  &   \textbf{21.49} &  \textbf{24.15} \\ \bottomrule
\end{tabular}
\end{table*}

The Dirichlet distribution is used to simulate the non-IID characteristics of CIFAR-10, CIFAR-100, and TinyImageNet datasets on federated learning. 
We denote the Dirichlet distribution by $Dir(N,\beta)$, where $N$ is the total number of local clients and $\beta$ represents the concentration parameter controlling the degree of non-IIDness of decentralized local data distributions. 
The probability $p_{k,j}$ is sampled from $Dir(N,\beta)$ and assigns the proportion of class $j$ in $k$-th client's dataset. 
With this non-IID distribution strategy with a low $\beta$ value, local clients will have a disparate class distribution from one another. The default values for $N$ and $\beta$ are set to 20 and 0.5, respectively.
In the case of FEMNIST, on the other hand, the writers of digits/characters in the data are randomly distributed to $N$ clients. $N$ is set to 20 for FEMNIST as well.
\looseness=-1

\cutparagraphup
\noindent \textbf{Implementation details.} 
The number of communication rounds is set to 100, with 1 epoch per round for all federated learning experiments. Half of the clients (i.e., $10=N/2$ clients) are randomly selected in each round for communication to make the federated setting more realistic. ResNet18~\cite{he2016deep} is used as the default backbone network following the literature in federated learning~\cite{li2021model,han2022fedx}. 
The learning rate ($\eta$), weight decay, and momentum for the SGD optimizer are set to 0.01, 1e-5, and 0.9, respectively. 
The batch size is set to 64. 
For knowledge distillation objectives, the temperature $\tau$ is set to 2.
Random crop, color jitter, and random horizontal flip are used for data augmentations in local model training. \looseness=-1

\cutparagraphup
\noindent\textbf{Model poisoning attack scenario.}
We divide a total of $N$ clients into benign and malicious clients with a ratio of 8:2 (i.e., attacker ratio = 20\%) as the default setting. 
For example, given $N$ to be 20, we assign four clients to play the adversarial role from a pool of 20 clients. Because we randomly select 10 out of 20 clients in every communication round, the number of malicious clients in each round may vary.
We consider two representative scenarios of an untargeted poisoning attack on the federated learning framework as follows: 
\looseness=-1

\cutparagraphup
\noindent\textbf{Scenario-1) No access to benign clients' information:} 
This scenario assumes that malicious clients cannot obtain any information about benign clients. 
We consider the label flipping attack as the model poisoning attack in this case, as it does not require prior knowledge of the training data or the benign clients' update information. For instance, in CIFAR-10, a class `dog' image may have a random label of `ship' or `horse'.
This false model update is sent to the central server after training the local network with randomly flipped noisy labels.  \looseness=-1

\cutparagraphup
\noindent\textbf{Scenario-2) Partial access to benign client information:} 
This scenario represents a more advanced attack in which malicious clients can access local updates from benign clients. 
Updates from malicious clients are computed using the statistics of benign updates.
We consider three variations:
\begin{enumerate}
\item \textsf{LIE} algorithm infers both the standard deviation and intensify factors from the updates of benign clients. Then, the perturbed noise is generated by multiplying the two to deceive the server-side defense~\cite{baruch2019little}.
\item \textsf{STAT-OPT} injects a static inverse unit vector into the central server, computed as the opposite direction of the noise from the mean of the benign clients' updates~\cite{fang2020local}.
\item \textsf{DYN-OPT} injects     dynamically perturbed noise, where the maximum distance from any other updates is bounded by the maximum distance between any two benign updates~\cite{shejwalkar2021manipulating}.
\end{enumerate} \looseness=-1
\cutsubsectionup
\subsection{Defense Performance Evaluation}
\cutparagraphup
\noindent\textbf{Baselines.} 
We have implemented six server-side defense strategies as baselines for untargeted attacks: (1) \textsf{No Defense} refers to the typical Federated Averaging (FedAvg) algorithm without any robust aggregation strategy. 
(2) \textsf{Median} is an aggregation algorithm that computes the median of each dimension of the updates rather than the average.
(3) \textsf{Trimmed Mean} is an aggregation algorithm that computes the mean of local updates by removing the largest and smallest values.
(4) \textsf{Norm Bound} is an algorithm that removes updates if the norm of the local update is above a certain threshold.
(5) \textsf{Multi-Krum} is an algorithm that iteratively selects local updates using the Krum method, which selects a single honest client by calculating the Euclidean distance between the client's update and the updates of its neighbors.
(6) \textsf{ResidualBase} is an algorithm that computes parameter confidence using the residuals of each parameter from the repeated mean. 

\cutparagraphup
\noindent\textbf{Evaluation.} 
All models are evaluated using the same attacks and experimental settings (e.g., client pool, ratio of attacker numbers, attack strategy, number of local epochs, optimizer, and communication rounds) to ensure a fair comparison.  
We report the top-1 classification accuracy on the test set, including the last and best accuracy. 
The reported accuracy (labeled \textsf{Last} and \textsf{Best}) is calculated by averaging the accuracy of the last and best five rounds, respectively. 

\begin{figure}[t!]
\centering
\includegraphics[width=0.45\textwidth]{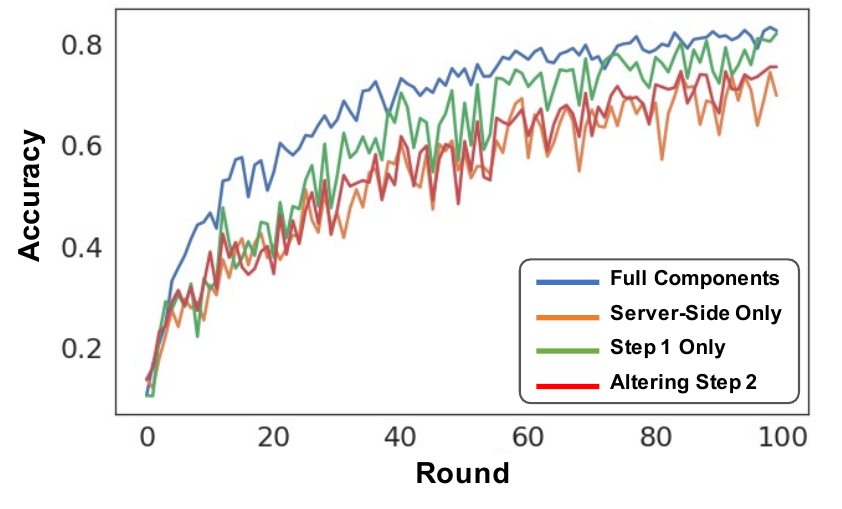}
\cutcationup
\caption{Performance comparison of ablations across communication rounds on CIFAR-10. Removing or altering any component results in a performance drop.}
\vspace{-3mm}
    \label{fig:ablation}
\end{figure}

\cutparagraphup
\noindent\textbf{Result.} 
Table~\ref{tab:main_result} compares the performance of defense algorithms against the label flipping attack described in Scenario 1.
We can see that \model{} consistently improves the performance when applied to existing server-side defense algorithms. 
Despite unfiltered malicious updates from failures in server-side defense or the absence of a defense strategy (i.e., \textsf{No Defense}), incorporating \model{} can lower the risk of performance degradation.

Table~\ref{tab:main_result2} reports the results against advanced attacks in Scenario 2, where malicious clients have access to information from benign clients. 
\model{} consistently outperforms the baselines across all cases. 
Server-side defenses alone can lead to removing updates of benign users during training, which can cause the performance to drop. 
For example, existing server-side defense strategies perform poorly in LIE cases, even compared to \textsf{No Defense}.
However, adding \model{} can alleviate this issue and improve performance, highlighting the effectiveness of our approach.

\cutsubsectionup
\subsection{Component Analyses}\cutparagraphup
\noindent \textbf{Ablation study.~} 
We have conducted an ablation study by removing and altering the key components.
We compare the following variations:  \looseness=-1
\begin{itemize}
\item Full Components: Include all components. 
\item Step 1 Only: Only using the local meta update component (Section~\ref{Sec:local_update}), with Step 2 removed.
\item Altering Step 2: Altering knowledge distillation (Section~\ref{Sec:global_update}) with a conventional alternative, which naively distills global knowledge without calibration.
\item Sever-Side Only: A model that uses only the server-side defense method without any attack-tolerant local updates. 
\end{itemize}
\looseness=-1

Figure~\ref{fig:ablation} confirms that the full model provides the best performance among the ablations. Due to space limitations, we report results when our method is used along with \textsf{Multi-Krum}.  
Ablations using Step 1 only shows a more robust performance than using the server-side defense strategy alone.
Altering Step 2 to a conventional knowledge distillation, on the other hand, degrades performance compared with entirely removing Step 2.
This is likely due to corrupted information in the global model, $F_\theta$, and demonstrates the need for a carefully designed attack-tolerant knowledge distillation.   

\begin{figure*}[t!]
\centering
\begin{subfigure}[t!]{0.27\textwidth}
\captionsetup{width=.99\columnwidth}
\includegraphics[width=0.99\columnwidth]{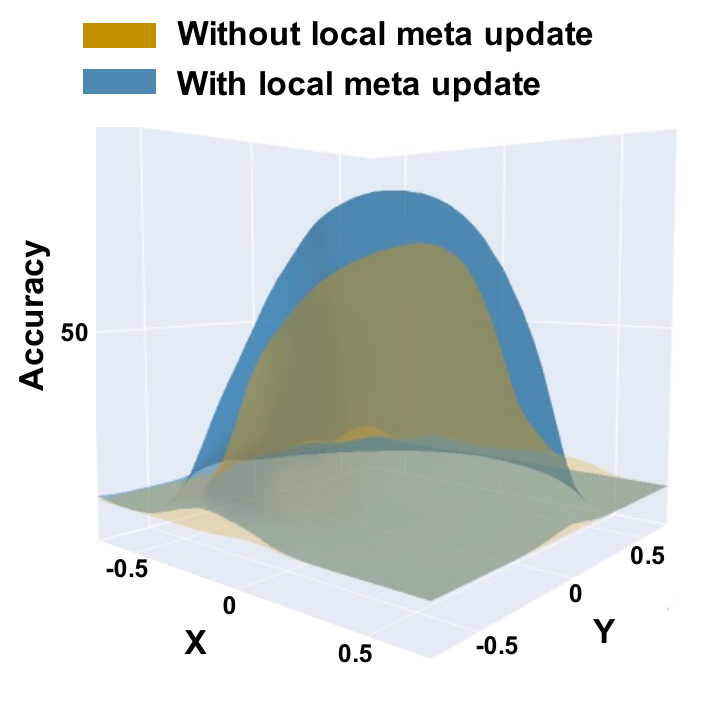}
    \cutcationup
    \vspace{-5.5mm}
    \caption{Visualization of accuracy surface}
    \label{fig:acc_surface}
\end{subfigure}
\hspace{1mm}
\begin{subfigure}[t!]{0.34\textwidth}
\captionsetup{width=.99\linewidth}
\includegraphics[width=0.99\columnwidth]{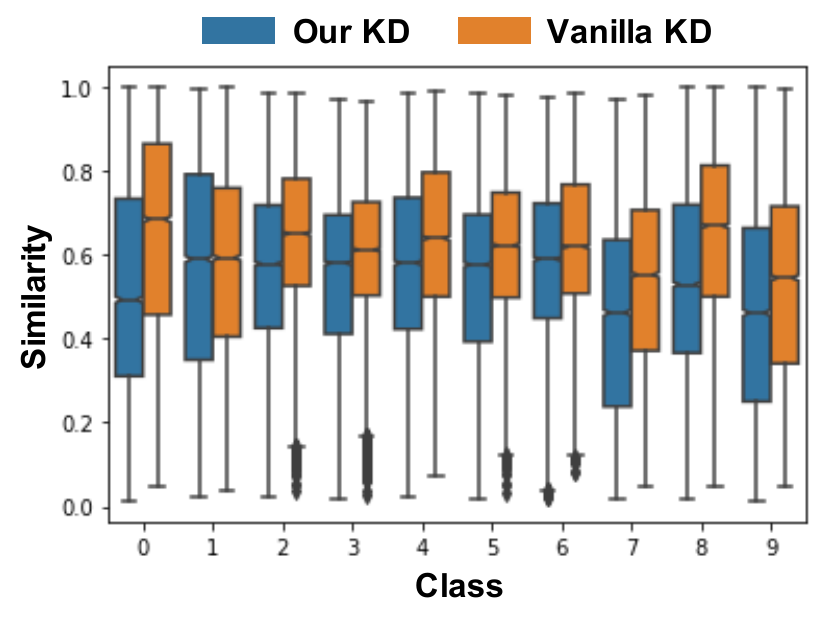}
    \cutcationup
    \caption{Similarity of local \& corrupted global~(\eqref{eq:corrupted_model})}
    \label{fig:pertur_class}
\end{subfigure}
\begin{subfigure}[t!]{0.34\textwidth}
\captionsetup{width=.99\linewidth}
\includegraphics[width=0.99\columnwidth]{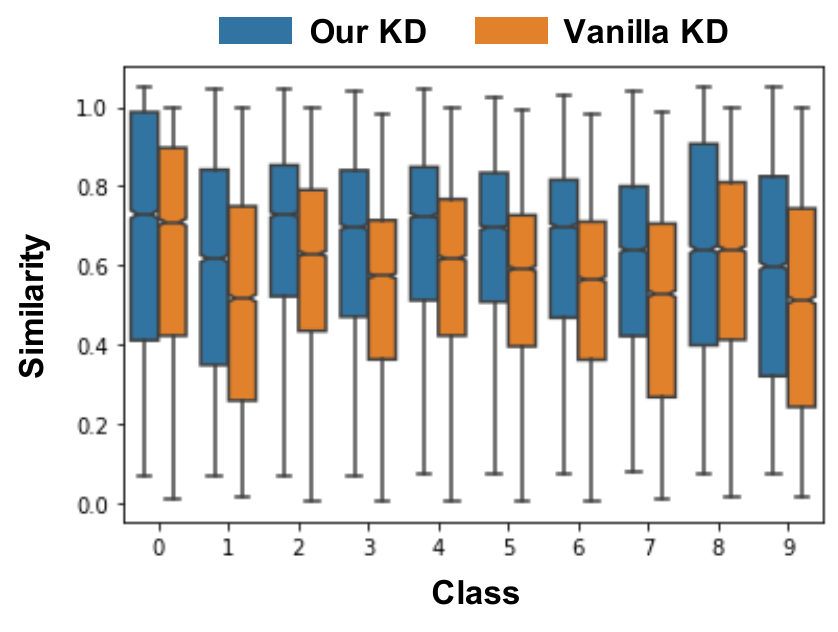}
    \cutcationup
    \caption{Similarity of local \& clean global~(\eqref{eq:clean_model})}
    \label{fig:clean_class}
\end{subfigure}
\cutcationup
\caption{Qualitative analyses on \model{}. (a) The accuracy surface of the global model with or without a local meta update after adding small perturbations to the model parameter. $\mathbf{X}$ and $\mathbf{Y}$ refer to two random directions for the perturbation. (b-c) The box plots the similarity between the local and global models (either corrupted or clean) with/without our proposed knowledge distillation. \looseness=-1}
\end{figure*}

\begin{table}[t!]
\caption{Performance comparison of the proposed knowledge distillation technique with existing global regularization approaches. The proposed knowledge distillation technique achieves the best accuracy. \looseness=-1}
\cutcationup
\centering
\begin{tabular}{l|cc|cc}

\toprule
\multirow{2}{*}{Global Knowledge Distillation} & \multicolumn{2}{c|}{No Defense} & \multicolumn{2}{c}{Multi-Krum} \\
 & ~~Last~~   &   ~~Best~~  &  ~~Last~~   &   ~~Best~~    \\  \midrule
 Full Components & \textbf{78.17} & \textbf{79.96} & \textbf{81.87} & \textbf{82.78} \\
 Local Meta Update + Scaffold & 74.81 & 78.52 & 80.32 & 80.57 \\
 Local Meta Update + FedProx & 74.67 & 78.01 & 80.83 & 82.01 \\
 Local Meta Update + Moon & 73.13 & 75.12 & 78.05 & 78.99  \\ \bottomrule
\end{tabular}
\label{tab:global_ablation}
\end{table}

\cutparagraphup
\noindent
\textbf{Comparison with alternative global regularization approaches.}
The proposed global knowledge distillation (Section~\ref{Sec:global_update}) may be replaced with existing global regularization techniques that don't consider malicious attacks. We have tested the three alternatives that replace the global knowledge distillation with existing methods FedProx~\cite{li2020fedprox}, Scaffold~\cite{karimireddy2020scaffold}, or Moon~\cite{li2021model} on top of the local meta update module. 
These existing methods also regularize the local model and address the local drift fallacy caused by the disagreement between local and global data distribution. 

Table~\ref{tab:global_ablation} compares our method with these alternatives over two different global aggregation strategies: simple averaging (i.e., No Defense) and Multi-Krum.
\model{} outperforms in both the last and best accuracy across both defense cases. 
When the global model is heavily perturbed (i.e., No Defense), the existing global knowledge distillation method experiences a significant drop in accuracy compared with when the global model is less perturbed (i.e., Multi-Krum). 
We also find that our proposed global knowledge distillation leads to smaller accuracy fluctuation across training rounds compared with other methods (see Figure~\ref{fig:global_ablation} in the Appendix).
These results demonstrate the critical role that our attack-tolerant global knowledge distillation plays.

\cutparagraphup
\noindent\textbf{Qualitative analyses.} 
We investigate how well \model{} discovers the attack-tolerant parameters for benign local models. 
We are inspired by the experiment to visualize the loss landscape in~\cite{visualloss} and follow it as follows. 
We add random direction perturbations to the model parameter (for example, for two random directions X and Y in Fig.~\ref{fig:acc_surface}) and examine how the accuracy changes.
Results in Fig.~\ref{fig:acc_surface} indicate that our local meta update (step 1) produces a higher and smoother accuracy surface compared with when this step is missing. 
Resistance to random parameter perturbations suggests that, when utilizing FedDefender, each benign client can find a solution with flat minima in the loss curve within the parameter space.

\begin{table}[t!]
\caption{Performance comparison over different experimental settings with varying hyper-parameters. The results demonstrate that \model{} consistently enhances the performance when it is applied on top of the existing aggregation strategy.}
\cutcationup
\begin{subtable}[t]{0.48\textwidth}
\centering

\begin{tabular}{c|cc|cc}
\toprule
\multirow{2}{*}{Client number ($N$)} & \multicolumn{2}{c|}{Multi-Krum} & \multicolumn{2}{c}{Multi-Krum+Ours} \\
 & ~~Last~~   &   ~~Best~~  &  ~~Last~   &   ~Best~~    \\  \midrule
 10 & 82.53 & 83.21  & ~\textbf{84.66} & \textbf{84.94} \\
 15 & 81.26 & 82.75 & ~\textbf{86.11} & \textbf{86.89} \\
 20 & 73.09 & 75.03 & ~\textbf{81.87} & \textbf{82.77} \\
 25 & 65.18 & 73.81 & ~\textbf{81.67} & \textbf{82.34} \\ \bottomrule
\end{tabular}
\label{tab:table1_b}
\caption{Effect of the number of clients}

\end{subtable}
\begin{subtable}[t]{0.48\textwidth}
\centering

\begin{tabular}{c|cc|cc}
\toprule
\multirow{2}{*}{Attacker ratio ($p_a$)} & \multicolumn{2}{c|}{Multi-Krum} & \multicolumn{2}{c}{Multi-Krum+Ours} \\
 &  ~~Last~~   &   ~~Best~~  &  ~~Last~   &   ~Best~~ \\  \midrule
10 & 74.11 & 76.72 & ~\textbf{80.75} & \textbf{82.74} \\ 
15 & 73.74 & 76.56 & ~\textbf{81.73} & \textbf{82.96} \\
20 & 73.09 & 75.03 & ~\textbf{81.87} & \textbf{82.77} \\
25 & 72.52 & 75.85 & ~\textbf{77.99} & \textbf{81.97} \\\bottomrule
\end{tabular}
\label{tab:table1_c}
\caption{Effect of the percentage of attackers}
\end{subtable}
\vspace{3mm}
\centering
\begin{subtable}[t]{0.48\textwidth}
\centering

\begin{tabular}{c|cc|cc}
\toprule
\multirow{2}{*}{Non-IIDness ($\beta$)} & \multicolumn{2}{c|}{Multi-Krum} & \multicolumn{2}{c}{Multi-Krum+Ours} \\
 &  ~~Last~~   &   ~~Best~~  & ~~Last~   &   ~Best~~   \\ \midrule
1.00                      & 79.49         & 80.74         & ~\textbf{84.62}         & \textbf{84.98}         \\ 
0.75                       &    79.65      &  80.56       & ~\textbf{83.70}         & \textbf{84.24}         \\
0.50                       & 73.09         & 75.03         & ~\textbf{81.87}         & \textbf{82.77}         \\
0.25                       &   66.66      &  68.71    & ~\textbf{71.68}         & \textbf{74.05}         \\ \bottomrule
\end{tabular}
\label{tab:table1_a}
\caption{Effect of the level of non-IIDness} \vspace{-7mm}
\end{subtable}
\label{tab:robustness}
\looseness=-1
\end{table}

Next, we examine how well the correct global knowledge is conveyed to the local model by \model{} under poisoning attacks.
Based on the experiment, we may exactly compute a hypothetical global model that is immune to attacks (i.e., a clean global model) by removing all adversarial clients in the global aggregation phase. 
Mathematically, the corrupted and clean global models in round $t$ are defined as follows:
\begin{align}
\theta_{\text{corrupted}}^{t+1} &= \theta^t + {{\sum_{k=1}^N \mathbbm{1}_{\{k \in S_{b} \cup S_{m}\}}} \cdot \Delta \theta_{k}^{t} \over {|S_b| + |S_m|}} \label{eq:corrupted_model}
\end{align}
\vspace{-3mm}
\begin{align}
\theta_{\text{clean}}^{t+1} &= \theta^t + {{\sum_{k=1}^N \mathbbm{1}_{\{k \in S_{b}\}}} \cdot \Delta \theta_{k}^{t} \over |S_b|}, \label{eq:clean_model} 
\end{align}
where $\mathcal{S}_b$ and $\mathcal{S}_m$ represent the sets of benign and malicious clients, respectively.
We compute the similarity between the local models with two global models: one is corrupted and the other is clean. 
Average cosine similarity between model predictions over the same dataset is used to evaluate the model similarity.

Figures~\ref{fig:pertur_class} and~\ref{fig:clean_class} show the relationship between local and global models for each class in CIFAR-10.
The results show that local models in \model{} have a substantially higher similarity to the clean global model (0.64) compared with the corrupted global model (0.53). 
In contrast, a vanilla knowledge distillation does not guarantee the same ability to filter out contaminated knowledge from the global model (clean: 0.55 vs corrupted: 0.61). 
This suggests that our method can effectively distill more correct and calibrated knowledge from the corrupted global model.

\cutparagraphup \noindent
\textbf{Robustness test.~} 
To test robustness, we consider different experimental settings and vary the key hyperparameters.
These include (a) the number of participating clients $N$, (b) the percentage of attackers that infiltrate the system $p_a$, and (c) the level of non-IIDness in the distributed local dataset, which is controlled by the beta parameter $\beta$ in Dirichlet distribution. A lower beta parameter leads to higher non-IIDness. Table~\ref{tab:robustness} shows the results for using Multi-Krum as a baseline aggregation strategy on CIFAR-10. 

The complexity of data training increases as we increase the number of clients, the percentage of attackers, and the level of non-IIDness. Irrespective of these changes, \model{} consistently demonstrates significant performance improvements.

\cutsectionup

\section{Conclusion}
This paper proposed \model{}, a client-side approach to improve existing server-side defense strategies against model poisoning attacks. 
We proposed two attack-tolerant training components for benign clients: (1) attack-tolerant local meta update and (2) attack-tolerant global knowledge distillation.
With these components, our method helps mitigate model poisoning attacks and produces more trustworthy results, even when server-side defenses fail to filter out malicious updates. 
As a result, our method has achieved a meaningful robustness improvement against various model poisoning attacks when used in conjunction with existing server-side defense strategies.
We hope that our technique can serve as a foundation for further research in client-side robust federated learning. \looseness=-1

\section*{Acknowledgement}
This research was supported by the Institute for Basic Science (IBS-R029-C2), Microsoft Research Asia, the Potential Individuals Global Training Program (2022-00155958), and the IITP grant (RS-2023-00216011) by the Ministry of Science and ICT in Korea.

\newpage
\balance
\bibliographystyle{ACM-Reference-Format}
\bibliography{main}


\begin{thebibliography}{49}

%%% the Web address.

\ifx \showCODEN    \undefined \def \showCODEN     #1{\unskip}     \fi
\ifx \showDOI      \undefined \def \showDOI       #1{#1}\fi
\ifx \showISBNx    \undefined \def \showISBNx     #1{\unskip}     \fi
\ifx \showISBNxiii \undefined \def \showISBNxiii  #1{\unskip}     \fi
\ifx \showISSN     \undefined \def \showISSN      #1{\unskip}     \fi
\ifx \showLCCN     \undefined \def \showLCCN      #1{\unskip}     \fi
\ifx \shownote     \undefined \def \shownote      #1{#1}          \fi
\ifx \showarticletitle \undefined \def \showarticletitle #1{#1}   \fi
\ifx \showURL      \undefined \def \showURL       {\relax}        \fi
% invisible to TeX
\providecommand\bibfield[2]{#2}
\providecommand\bibinfo[2]{#2}
\providecommand\natexlab[1]{#1}
\providecommand\showeprint[2][]{arXiv:#2}

\bibitem[\protect\citeauthoryear{Awan, Luo, and Li}{Awan et~al\mbox{.}}{2021}]%
        {awan2021contra}
\bibfield{author}{\bibinfo{person}{Sana Awan}, \bibinfo{person}{Bo Luo}, {and}
  \bibinfo{person}{Fengjun Li}.} \bibinfo{year}{2021}\natexlab{}.
\newblock \showarticletitle{Contra: Defending against poisoning attacks in
  federated learning}. In \bibinfo{booktitle}{\emph{Proceedings of ESORICS}}.
  Springer, \bibinfo{pages}{455--475}.
\newblock


\bibitem[\protect\citeauthoryear{Baldock, Maennel, and Neyshabur}{Baldock
  et~al\mbox{.}}{2021}]%
        {baldock2021deep}
\bibfield{author}{\bibinfo{person}{Robert Baldock}, \bibinfo{person}{Hartmut
  Maennel}, {and} \bibinfo{person}{Behnam Neyshabur}.}
  \bibinfo{year}{2021}\natexlab{}.
\newblock \showarticletitle{Deep learning through the lens of example
  difficulty}. In \bibinfo{booktitle}{\emph{Advances in NeurIPS}},
  Vol.~\bibinfo{volume}{34}. \bibinfo{pages}{10876--10889}.
\newblock


\bibitem[\protect\citeauthoryear{Baruch, Baruch, and Goldberg}{Baruch
  et~al\mbox{.}}{2019}]%
        {baruch2019little}
\bibfield{author}{\bibinfo{person}{Gilad Baruch}, \bibinfo{person}{Moran
  Baruch}, {and} \bibinfo{person}{Yoav Goldberg}.}
  \bibinfo{year}{2019}\natexlab{}.
\newblock \showarticletitle{A little is enough: Circumventing defenses for
  distributed learning}. In \bibinfo{booktitle}{\emph{Advances in NeurIPS}},
  Vol.~\bibinfo{volume}{32}.
\newblock


\bibitem[\protect\citeauthoryear{Blanchard, El~Mhamdi, Guerraoui, and
  Stainer}{Blanchard et~al\mbox{.}}{2017}]%
        {blanchard2017machine}
\bibfield{author}{\bibinfo{person}{Peva Blanchard}, \bibinfo{person}{El~Mahdi
  El~Mhamdi}, \bibinfo{person}{Rachid Guerraoui}, {and} \bibinfo{person}{Julien
  Stainer}.} \bibinfo{year}{2017}\natexlab{}.
\newblock \showarticletitle{Machine learning with adversaries: Byzantine
  tolerant gradient descent}. In \bibinfo{booktitle}{\emph{Advances in
  NeurIPS}}, Vol.~\bibinfo{volume}{30}.
\newblock


\bibitem[\protect\citeauthoryear{Bonawitz, Eichner, Grieskamp, Huba, Ingerman,
  Ivanov, Kiddon, Kone{\v{c}}n{\`y}, Mazzocchi, McMahan,
  et~al\mbox{.}}{Bonawitz et~al\mbox{.}}{2019}]%
        {bonawitz2019towards}
\bibfield{author}{\bibinfo{person}{Keith Bonawitz}, \bibinfo{person}{Hubert
  Eichner}, \bibinfo{person}{Wolfgang Grieskamp}, \bibinfo{person}{Dzmitry
  Huba}, \bibinfo{person}{Alex Ingerman}, \bibinfo{person}{Vladimir Ivanov},
  \bibinfo{person}{Chloe Kiddon}, \bibinfo{person}{Jakub Kone{\v{c}}n{\`y}},
  \bibinfo{person}{Stefano Mazzocchi}, \bibinfo{person}{Brendan McMahan},
  {et~al\mbox{.}}} \bibinfo{year}{2019}\natexlab{}.
\newblock \showarticletitle{Towards federated learning at scale: System
  design}. In \bibinfo{booktitle}{\emph{Proceedings of MLSys}},
  Vol.~\bibinfo{volume}{1}. \bibinfo{pages}{374--388}.
\newblock


\bibitem[\protect\citeauthoryear{Caldas, Duddu, Wu, Li, Kone{\v{c}}n{\`y},
  McMahan, Smith, and Talwalkar}{Caldas et~al\mbox{.}}{2018}]%
        {caldas2018leaf}
\bibfield{author}{\bibinfo{person}{Sebastian Caldas}, \bibinfo{person}{Sai
  Meher~Karthik Duddu}, \bibinfo{person}{Peter Wu}, \bibinfo{person}{Tian Li},
  \bibinfo{person}{Jakub Kone{\v{c}}n{\`y}}, \bibinfo{person}{H~Brendan
  McMahan}, \bibinfo{person}{Virginia Smith}, {and} \bibinfo{person}{Ameet
  Talwalkar}.} \bibinfo{year}{2018}\natexlab{}.
\newblock \showarticletitle{Leaf: A benchmark for federated settings}.
\newblock \bibinfo{journal}{\emph{arXiv preprint arXiv:1812.01097}}
  (\bibinfo{year}{2018}).
\newblock


\bibitem[\protect\citeauthoryear{Chen, Pan, Monga, Bengio, and Jozefowicz}{Chen
  et~al\mbox{.}}{2016}]%
        {chen2016revisiting}
\bibfield{author}{\bibinfo{person}{Jianmin Chen}, \bibinfo{person}{Xinghao
  Pan}, \bibinfo{person}{Rajat Monga}, \bibinfo{person}{Samy Bengio}, {and}
  \bibinfo{person}{Rafal Jozefowicz}.} \bibinfo{year}{2016}\natexlab{}.
\newblock \showarticletitle{Revisiting distributed synchronous SGD}.
\newblock \bibinfo{journal}{\emph{arXiv preprint arXiv:1604.00981}}
  (\bibinfo{year}{2016}).
\newblock


\bibitem[\protect\citeauthoryear{Chen, Gui, Lin, Gan, and Wu}{Chen
  et~al\mbox{.}}{2022}]%
        {chen2022federated}
\bibfield{author}{\bibinfo{person}{Yao Chen}, \bibinfo{person}{Yijie Gui},
  \bibinfo{person}{Hong Lin}, \bibinfo{person}{Wensheng Gan}, {and}
  \bibinfo{person}{Yongdong Wu}.} \bibinfo{year}{2022}\natexlab{}.
\newblock \showarticletitle{Federated Learning Attacks and Defenses: A Survey}.
\newblock \bibinfo{journal}{\emph{arXiv preprint arXiv:2211.14952}}
  (\bibinfo{year}{2022}).
\newblock


\bibitem[\protect\citeauthoryear{Fang, Cao, Jia, and Gong}{Fang
  et~al\mbox{.}}{2020}]%
        {fang2020local}
\bibfield{author}{\bibinfo{person}{Minghong Fang}, \bibinfo{person}{Xiaoyu
  Cao}, \bibinfo{person}{Jinyuan Jia}, {and} \bibinfo{person}{Neil Gong}.}
  \bibinfo{year}{2020}\natexlab{}.
\newblock \showarticletitle{Local model poisoning attacks to
  $\{$Byzantine-Robust$\}$ federated learning}. In
  \bibinfo{booktitle}{\emph{Proceedings of USENIX Security}}.
  \bibinfo{pages}{1605--1622}.
\newblock


\bibitem[\protect\citeauthoryear{Finn, Abbeel, and Levine}{Finn
  et~al\mbox{.}}{2017}]%
        {finn2017model}
\bibfield{author}{\bibinfo{person}{Chelsea Finn}, \bibinfo{person}{Pieter
  Abbeel}, {and} \bibinfo{person}{Sergey Levine}.}
  \bibinfo{year}{2017}\natexlab{}.
\newblock \showarticletitle{Model-agnostic meta-learning for fast adaptation of
  deep networks}. In \bibinfo{booktitle}{\emph{Proceedings of ICML}}.
  \bibinfo{pages}{1126--1135}.
\newblock


\bibitem[\protect\citeauthoryear{Fu, Xie, Li, and Chen}{Fu
  et~al\mbox{.}}{2019}]%
        {fu2019attack}
\bibfield{author}{\bibinfo{person}{Shuhao Fu}, \bibinfo{person}{Chulin Xie},
  \bibinfo{person}{Bo Li}, {and} \bibinfo{person}{Qifeng Chen}.}
  \bibinfo{year}{2019}\natexlab{}.
\newblock \showarticletitle{Attack-resistant federated learning with
  residual-based reweighting}.
\newblock \bibinfo{journal}{\emph{arXiv preprint arXiv:1912.11464}}
  (\bibinfo{year}{2019}).
\newblock


\bibitem[\protect\citeauthoryear{Fung, Yoon, and Beschastnikh}{Fung
  et~al\mbox{.}}{2020}]%
        {fung2020limitations}
\bibfield{author}{\bibinfo{person}{Clement Fung}, \bibinfo{person}{Chris~JM
  Yoon}, {and} \bibinfo{person}{Ivan Beschastnikh}.}
  \bibinfo{year}{2020}\natexlab{}.
\newblock \showarticletitle{The Limitations of Federated Learning in Sybil
  Settings}. In \bibinfo{booktitle}{\emph{Proceedings of RAID}}.
\newblock


\bibitem[\protect\citeauthoryear{Gu, Dolan-Gavitt, and Garg}{Gu
  et~al\mbox{.}}{2017}]%
        {gu2017badnets}
\bibfield{author}{\bibinfo{person}{Tianyu Gu}, \bibinfo{person}{Brendan
  Dolan-Gavitt}, {and} \bibinfo{person}{Siddharth Garg}.}
  \bibinfo{year}{2017}\natexlab{}.
\newblock \showarticletitle{Badnets: Identifying vulnerabilities in the machine
  learning model supply chain}.
\newblock \bibinfo{journal}{\emph{arXiv preprint arXiv:1708.06733}}
  (\bibinfo{year}{2017}).
\newblock


\bibitem[\protect\citeauthoryear{Han, Park, Park, Kim, and Cha}{Han
  et~al\mbox{.}}{2020}]%
        {han2020mitigating}
\bibfield{author}{\bibinfo{person}{Sungwon Han}, \bibinfo{person}{Sungwon
  Park}, \bibinfo{person}{Sungkyu Park}, \bibinfo{person}{Sundong Kim}, {and}
  \bibinfo{person}{Meeyoung Cha}.} \bibinfo{year}{2020}\natexlab{}.
\newblock \showarticletitle{Mitigating embedding and class assignment mismatch
  in unsupervised image classification}. In
  \bibinfo{booktitle}{\emph{Proceedings of ECCV}}. Springer,
  \bibinfo{pages}{768--784}.
\newblock


\bibitem[\protect\citeauthoryear{Han, Park, Wu, Kim, Wu, Xie, and Cha}{Han
  et~al\mbox{.}}{2022}]%
        {han2022fedx}
\bibfield{author}{\bibinfo{person}{Sungwon Han}, \bibinfo{person}{Sungwon
  Park}, \bibinfo{person}{Fangzhao Wu}, \bibinfo{person}{Sundong Kim},
  \bibinfo{person}{Chuhan Wu}, \bibinfo{person}{Xing Xie}, {and}
  \bibinfo{person}{Meeyoung Cha}.} \bibinfo{year}{2022}\natexlab{}.
\newblock \showarticletitle{FedX: Unsupervised Federated Learning with Cross
  Knowledge Distillation}. In \bibinfo{booktitle}{\emph{Proceedings of ECCV}}.
  \bibinfo{pages}{691--707}.
\newblock


\bibitem[\protect\citeauthoryear{He, Zhang, Ren, and Sun}{He
  et~al\mbox{.}}{2016}]%
        {he2016deep}
\bibfield{author}{\bibinfo{person}{Kaiming He}, \bibinfo{person}{Xiangyu
  Zhang}, \bibinfo{person}{Shaoqing Ren}, {and} \bibinfo{person}{Jian Sun}.}
  \bibinfo{year}{2016}\natexlab{}.
\newblock \showarticletitle{Deep residual learning for image recognition}. In
  \bibinfo{booktitle}{\emph{Proceedings of CVPR}}. \bibinfo{pages}{770--778}.
\newblock


\bibitem[\protect\citeauthoryear{Hinton, Vinyals, Dean, et~al\mbox{.}}{Hinton
  et~al\mbox{.}}{2015}]%
        {hinton2015distilling}
\bibfield{author}{\bibinfo{person}{Geoffrey Hinton}, \bibinfo{person}{Oriol
  Vinyals}, \bibinfo{person}{Jeff Dean}, {et~al\mbox{.}}}
  \bibinfo{year}{2015}\natexlab{}.
\newblock \showarticletitle{Distilling the knowledge in a neural network}.
\newblock \bibinfo{journal}{\emph{arXiv preprint arXiv:1503.02531}}
  \bibinfo{volume}{2}, \bibinfo{number}{7} (\bibinfo{year}{2015}).
\newblock


\bibitem[\protect\citeauthoryear{Karimireddy, He, and Jaggi}{Karimireddy
  et~al\mbox{.}}{[n.\,d.]}]%
        {karimireddybyzantine}
\bibfield{author}{\bibinfo{person}{Sai~Praneeth Karimireddy},
  \bibinfo{person}{Lie He}, {and} \bibinfo{person}{Martin Jaggi}.}
  \bibinfo{year}{[n.\,d.]}\natexlab{}.
\newblock \showarticletitle{Byzantine-Robust Learning on Heterogeneous Datasets
  via Bucketing}. In \bibinfo{booktitle}{\emph{Proceedings of ICLR}}.
\newblock


\bibitem[\protect\citeauthoryear{Karimireddy, Kale, Mohri, Reddi, Stich, and
  Suresh}{Karimireddy et~al\mbox{.}}{2020}]%
        {karimireddy2020scaffold}
\bibfield{author}{\bibinfo{person}{Sai~Praneeth Karimireddy},
  \bibinfo{person}{Satyen Kale}, \bibinfo{person}{Mehryar Mohri},
  \bibinfo{person}{Sashank Reddi}, \bibinfo{person}{Sebastian Stich}, {and}
  \bibinfo{person}{Ananda~Theertha Suresh}.} \bibinfo{year}{2020}\natexlab{}.
\newblock \showarticletitle{Scaffold: Stochastic controlled averaging for
  federated learning}. In \bibinfo{booktitle}{\emph{Proceedings of ICML}}.
  \bibinfo{pages}{5132--5143}.
\newblock


\bibitem[\protect\citeauthoryear{Krizhevsky, Hinton, et~al\mbox{.}}{Krizhevsky
  et~al\mbox{.}}{2009}]%
        {krizhevsky2009learning}
\bibfield{author}{\bibinfo{person}{Alex Krizhevsky}, \bibinfo{person}{Geoffrey
  Hinton}, {et~al\mbox{.}}} \bibinfo{year}{2009}\natexlab{}.
\newblock \showarticletitle{Learning multiple layers of features from tiny
  images}.
\newblock  (\bibinfo{year}{2009}).
\newblock


\bibitem[\protect\citeauthoryear{Kundu, Sun, Fu, Pedram, and Beerel}{Kundu
  et~al\mbox{.}}{2021}]%
        {kundu2021analyzing}
\bibfield{author}{\bibinfo{person}{Souvik Kundu}, \bibinfo{person}{Qirui Sun},
  \bibinfo{person}{Yao Fu}, \bibinfo{person}{Massoud Pedram}, {and}
  \bibinfo{person}{Peter Beerel}.} \bibinfo{year}{2021}\natexlab{}.
\newblock \showarticletitle{Analyzing the confidentiality of undistillable
  teachers in knowledge distillation}. In \bibinfo{booktitle}{\emph{Advances in
  NeurIPS}}, Vol.~\bibinfo{volume}{34}. \bibinfo{pages}{9181--9192}.
\newblock


\bibitem[\protect\citeauthoryear{Li, Xu, Taylor, Studer, and Goldstein}{Li
  et~al\mbox{.}}{2018}]%
        {visualloss}
\bibfield{author}{\bibinfo{person}{Hao Li}, \bibinfo{person}{Zheng Xu},
  \bibinfo{person}{Gavin Taylor}, \bibinfo{person}{Christoph Studer}, {and}
  \bibinfo{person}{Tom Goldstein}.} \bibinfo{year}{2018}\natexlab{}.
\newblock \showarticletitle{Visualizing the Loss Landscape of Neural Nets}. In
  \bibinfo{booktitle}{\emph{Advances in NeurIPS}}.
\newblock


\bibitem[\protect\citeauthoryear{Li, Wong, Zhao, and Kankanhalli}{Li
  et~al\mbox{.}}{2019}]%
        {li2019learning}
\bibfield{author}{\bibinfo{person}{Junnan Li}, \bibinfo{person}{Yongkang Wong},
  \bibinfo{person}{Qi Zhao}, {and} \bibinfo{person}{Mohan~S Kankanhalli}.}
  \bibinfo{year}{2019}\natexlab{}.
\newblock \showarticletitle{Learning to learn from noisy labeled data}. In
  \bibinfo{booktitle}{\emph{Proceedings of CVPR}}. \bibinfo{pages}{5051--5059}.
\newblock


\bibitem[\protect\citeauthoryear{Li, He, and Song}{Li et~al\mbox{.}}{2021}]%
        {li2021model}
\bibfield{author}{\bibinfo{person}{Qinbin Li}, \bibinfo{person}{Bingsheng He},
  {and} \bibinfo{person}{Dawn Song}.} \bibinfo{year}{2021}\natexlab{}.
\newblock \showarticletitle{Model-contrastive federated learning}. In
  \bibinfo{booktitle}{\emph{Proceedings of CVPR}}.
  \bibinfo{pages}{10713--10722}.
\newblock


\bibitem[\protect\citeauthoryear{Li, Sahu, Zaheer, Sanjabi, Talwalkar, and
  Smith}{Li et~al\mbox{.}}{2020}]%
        {li2020fedprox}
\bibfield{author}{\bibinfo{person}{Tian Li}, \bibinfo{person}{Anit~Kumar Sahu},
  \bibinfo{person}{Manzil Zaheer}, \bibinfo{person}{Maziar Sanjabi},
  \bibinfo{person}{Ameet Talwalkar}, {and} \bibinfo{person}{Virginia Smith}.}
  \bibinfo{year}{2020}\natexlab{}.
\newblock \showarticletitle{Federated optimization in heterogeneous networks}.
\newblock \bibinfo{journal}{\emph{Proceedings of Machine Learning and Systems}}
   \bibinfo{volume}{2} (\bibinfo{year}{2020}), \bibinfo{pages}{429--450}.
\newblock


\bibitem[\protect\citeauthoryear{Luo, Li, Wang, Huang, and Tassiulas}{Luo
  et~al\mbox{.}}{2021}]%
        {luo2021cost}
\bibfield{author}{\bibinfo{person}{Bing Luo}, \bibinfo{person}{Xiang Li},
  \bibinfo{person}{Shiqiang Wang}, \bibinfo{person}{Jianwei Huang}, {and}
  \bibinfo{person}{Leandros Tassiulas}.} \bibinfo{year}{2021}\natexlab{}.
\newblock \showarticletitle{Cost-effective federated learning in mobile edge
  networks}.
\newblock \bibinfo{journal}{\emph{IEEE Journal on Selected Areas in
  Communications}} \bibinfo{volume}{39}, \bibinfo{number}{12}
  (\bibinfo{year}{2021}), \bibinfo{pages}{3606--3621}.
\newblock


\bibitem[\protect\citeauthoryear{Lyu, Yu, and Yang}{Lyu et~al\mbox{.}}{2020}]%
        {lyu2020threats}
\bibfield{author}{\bibinfo{person}{Lingjuan Lyu}, \bibinfo{person}{Han Yu},
  {and} \bibinfo{person}{Qiang Yang}.} \bibinfo{year}{2020}\natexlab{}.
\newblock \showarticletitle{Threats to federated learning: A survey}.
\newblock \bibinfo{journal}{\emph{arXiv preprint arXiv:2003.02133}}
  (\bibinfo{year}{2020}).
\newblock


\bibitem[\protect\citeauthoryear{McMahan, Moore, Ramage, Hampson, and
  y~Arcas}{McMahan et~al\mbox{.}}{2017}]%
        {mcmahan2017communication}
\bibfield{author}{\bibinfo{person}{Brendan McMahan}, \bibinfo{person}{Eider
  Moore}, \bibinfo{person}{Daniel Ramage}, \bibinfo{person}{Seth Hampson},
  {and} \bibinfo{person}{Blaise~Aguera y Arcas}.}
  \bibinfo{year}{2017}\natexlab{}.
\newblock \showarticletitle{Communication-efficient learning of deep networks
  from decentralized data}. In \bibinfo{booktitle}{\emph{Proceedings of
  AISTATS}}. \bibinfo{pages}{1273--1282}.
\newblock


\bibitem[\protect\citeauthoryear{Park, Han, Kim, Kim, Park, Hong, and Cha}{Park
  et~al\mbox{.}}{2021}]%
        {park2021improving}
\bibfield{author}{\bibinfo{person}{Sungwon Park}, \bibinfo{person}{Sungwon
  Han}, \bibinfo{person}{Sundong Kim}, \bibinfo{person}{Danu Kim},
  \bibinfo{person}{Sungkyu Park}, \bibinfo{person}{Seunghoon Hong}, {and}
  \bibinfo{person}{Meeyoung Cha}.} \bibinfo{year}{2021}\natexlab{}.
\newblock \showarticletitle{Improving unsupervised image clustering with robust
  learning}. In \bibinfo{booktitle}{\emph{Proceedings of CVPR}}.
  \bibinfo{pages}{12278--12287}.
\newblock


\bibitem[\protect\citeauthoryear{Park, Kim, and Cha}{Park
  et~al\mbox{.}}{2022}]%
        {park2022knowledge}
\bibfield{author}{\bibinfo{person}{Sungwon Park}, \bibinfo{person}{Sundong
  Kim}, {and} \bibinfo{person}{Meeyoung Cha}.} \bibinfo{year}{2022}\natexlab{}.
\newblock \showarticletitle{Knowledge sharing via domain adaptation in customs
  fraud detection}. In \bibinfo{booktitle}{\emph{Proceedings of AAAI}},
  Vol.~\bibinfo{volume}{36}. \bibinfo{pages}{12062--12070}.
\newblock


\bibitem[\protect\citeauthoryear{Pillutla, Kakade, and Harchaoui}{Pillutla
  et~al\mbox{.}}{2022}]%
        {pillutla2022robust}
\bibfield{author}{\bibinfo{person}{Krishna Pillutla}, \bibinfo{person}{Sham~M
  Kakade}, {and} \bibinfo{person}{Zaid Harchaoui}.}
  \bibinfo{year}{2022}\natexlab{}.
\newblock \showarticletitle{Robust aggregation for federated learning}.
\newblock \bibinfo{journal}{\emph{IEEE Transactions on Signal Processing}}
  \bibinfo{volume}{70} (\bibinfo{year}{2022}), \bibinfo{pages}{1142--1154}.
\newblock


\bibitem[\protect\citeauthoryear{Ravi and Larochelle}{Ravi and
  Larochelle}{2017}]%
        {ravi2017optimization}
\bibfield{author}{\bibinfo{person}{Sachin Ravi} {and} \bibinfo{person}{Hugo
  Larochelle}.} \bibinfo{year}{2017}\natexlab{}.
\newblock \showarticletitle{Optimization as a model for few-shot learning}. In
  \bibinfo{booktitle}{\emph{Proceedings of ICLR}}.
\newblock


\bibitem[\protect\citeauthoryear{Rieke, Hancox, Li, Milletari, Roth,
  Albarqouni, Bakas, Galtier, Landman, Maier-Hein, et~al\mbox{.}}{Rieke
  et~al\mbox{.}}{2020}]%
        {rieke2020future}
\bibfield{author}{\bibinfo{person}{Nicola Rieke}, \bibinfo{person}{Jonny
  Hancox}, \bibinfo{person}{Wenqi Li}, \bibinfo{person}{Fausto Milletari},
  \bibinfo{person}{Holger~R Roth}, \bibinfo{person}{Shadi Albarqouni},
  \bibinfo{person}{Spyridon Bakas}, \bibinfo{person}{Mathieu~N Galtier},
  \bibinfo{person}{Bennett~A Landman}, \bibinfo{person}{Klaus Maier-Hein},
  {et~al\mbox{.}}} \bibinfo{year}{2020}\natexlab{}.
\newblock \showarticletitle{The future of digital health with federated
  learning}.
\newblock \bibinfo{journal}{\emph{NPJ digital medicine}} \bibinfo{volume}{3},
  \bibinfo{number}{1} (\bibinfo{year}{2020}), \bibinfo{pages}{1--7}.
\newblock


\bibitem[\protect\citeauthoryear{Shafahi, Huang, Najibi, Suciu, Studer,
  Dumitras, and Goldstein}{Shafahi et~al\mbox{.}}{2018}]%
        {shafahi2018poison}
\bibfield{author}{\bibinfo{person}{Ali Shafahi}, \bibinfo{person}{W~Ronny
  Huang}, \bibinfo{person}{Mahyar Najibi}, \bibinfo{person}{Octavian Suciu},
  \bibinfo{person}{Christoph Studer}, \bibinfo{person}{Tudor Dumitras}, {and}
  \bibinfo{person}{Tom Goldstein}.} \bibinfo{year}{2018}\natexlab{}.
\newblock \showarticletitle{Poison frogs! targeted clean-label poisoning
  attacks on neural networks}. In \bibinfo{booktitle}{\emph{Advances in
  NeurIPS}}, Vol.~\bibinfo{volume}{31}.
\newblock


\bibitem[\protect\citeauthoryear{Shejwalkar and Houmansadr}{Shejwalkar and
  Houmansadr}{2021}]%
        {shejwalkar2021manipulating}
\bibfield{author}{\bibinfo{person}{Virat Shejwalkar} {and}
  \bibinfo{person}{Amir Houmansadr}.} \bibinfo{year}{2021}\natexlab{}.
\newblock \showarticletitle{Manipulating the byzantine: Optimizing model
  poisoning attacks and defenses for federated learning}. In
  \bibinfo{booktitle}{\emph{Proceedings of NDSS Symposium}}.
\newblock


\bibitem[\protect\citeauthoryear{Shejwalkar, Houmansadr, Kairouz, and
  Ramage}{Shejwalkar et~al\mbox{.}}{2022}]%
        {shejwalkar2022back}
\bibfield{author}{\bibinfo{person}{Virat Shejwalkar}, \bibinfo{person}{Amir
  Houmansadr}, \bibinfo{person}{Peter Kairouz}, {and} \bibinfo{person}{Daniel
  Ramage}.} \bibinfo{year}{2022}\natexlab{}.
\newblock \showarticletitle{Back to the drawing board: A critical evaluation of
  poisoning attacks on production federated learning}. In
  \bibinfo{booktitle}{\emph{Proceedings of the IEEE Symposium on Security and
  Privacy}}. \bibinfo{pages}{1354--1371}.
\newblock


\bibitem[\protect\citeauthoryear{Sikandar, Waheed, Tahir, Malik, and
  Rafique}{Sikandar et~al\mbox{.}}{2023}]%
        {sikandar2023detailed}
\bibfield{author}{\bibinfo{person}{Hira~Shahzadi Sikandar},
  \bibinfo{person}{Huda Waheed}, \bibinfo{person}{Sibgha Tahir},
  \bibinfo{person}{Saif~UR Malik}, {and} \bibinfo{person}{Waqas Rafique}.}
  \bibinfo{year}{2023}\natexlab{}.
\newblock \showarticletitle{A Detailed Survey on Federated Learning Attacks and
  Defenses}.
\newblock \bibinfo{journal}{\emph{Electronics}} \bibinfo{volume}{12},
  \bibinfo{number}{2} (\bibinfo{year}{2023}), \bibinfo{pages}{260}.
\newblock


\bibitem[\protect\citeauthoryear{Stephenson, Padhy, Ganesh, Hui, Tang, and
  Chung}{Stephenson et~al\mbox{.}}{2021}]%
        {stephenson2021geometry}
\bibfield{author}{\bibinfo{person}{Cory Stephenson},
  \bibinfo{person}{Suchismita Padhy}, \bibinfo{person}{Abhinav Ganesh},
  \bibinfo{person}{Yue Hui}, \bibinfo{person}{Hanlin Tang}, {and}
  \bibinfo{person}{SueYeon Chung}.} \bibinfo{year}{2021}\natexlab{}.
\newblock \showarticletitle{On the geometry of generalization and memorization
  in deep neural networks}. In \bibinfo{booktitle}{\emph{Proceedings of ICLR}}.
\newblock


\bibitem[\protect\citeauthoryear{Sun, Li, DiValentin, Hassanzadeh, Chen, and
  Li}{Sun et~al\mbox{.}}{2021}]%
        {sun2021fl}
\bibfield{author}{\bibinfo{person}{Jingwei Sun}, \bibinfo{person}{Ang Li},
  \bibinfo{person}{Louis DiValentin}, \bibinfo{person}{Amin Hassanzadeh},
  \bibinfo{person}{Yiran Chen}, {and} \bibinfo{person}{Hai Li}.}
  \bibinfo{year}{2021}\natexlab{}.
\newblock \showarticletitle{Fl-wbc: Enhancing robustness against model
  poisoning attacks in federated learning from a client perspective}.
\newblock \bibinfo{journal}{\emph{Advances in NeurIPS}}  \bibinfo{volume}{34}
  (\bibinfo{year}{2021}), \bibinfo{pages}{12613--12624}.
\newblock


\bibitem[\protect\citeauthoryear{Sun, Kairouz, Suresh, and McMahan}{Sun
  et~al\mbox{.}}{2019}]%
        {sun2019can}
\bibfield{author}{\bibinfo{person}{Ziteng Sun}, \bibinfo{person}{Peter
  Kairouz}, \bibinfo{person}{Ananda~Theertha Suresh}, {and}
  \bibinfo{person}{H~Brendan McMahan}.} \bibinfo{year}{2019}\natexlab{}.
\newblock \showarticletitle{Can you really backdoor federated learning?}
\newblock \bibinfo{journal}{\emph{arXiv preprint arXiv:1911.07963}}
  (\bibinfo{year}{2019}).
\newblock


\bibitem[\protect\citeauthoryear{Wang, Charles, Xu, Joshi, McMahan,
  Al-Shedivat, Andrew, Avestimehr, Daly, Data, et~al\mbox{.}}{Wang
  et~al\mbox{.}}{2021}]%
        {wang2021field}
\bibfield{author}{\bibinfo{person}{Jianyu Wang}, \bibinfo{person}{Zachary
  Charles}, \bibinfo{person}{Zheng Xu}, \bibinfo{person}{Gauri Joshi},
  \bibinfo{person}{H~Brendan McMahan}, \bibinfo{person}{Maruan Al-Shedivat},
  \bibinfo{person}{Galen Andrew}, \bibinfo{person}{Salman Avestimehr},
  \bibinfo{person}{Katharine Daly}, \bibinfo{person}{Deepesh Data},
  {et~al\mbox{.}}} \bibinfo{year}{2021}\natexlab{}.
\newblock \showarticletitle{A field guide to federated optimization}.
\newblock \bibinfo{journal}{\emph{arXiv preprint arXiv:2107.06917}}
  (\bibinfo{year}{2021}).
\newblock


\bibitem[\protect\citeauthoryear{Wu, Wu, Lyu, Huang, and Xie}{Wu
  et~al\mbox{.}}{2022a}]%
        {wu2022communication}
\bibfield{author}{\bibinfo{person}{Chuhan Wu}, \bibinfo{person}{Fangzhao Wu},
  \bibinfo{person}{Lingjuan Lyu}, \bibinfo{person}{Yongfeng Huang}, {and}
  \bibinfo{person}{Xing Xie}.} \bibinfo{year}{2022}\natexlab{a}.
\newblock \showarticletitle{Communication-efficient federated learning via
  knowledge distillation}.
\newblock \bibinfo{journal}{\emph{Nature communications}} \bibinfo{volume}{13},
  \bibinfo{number}{1} (\bibinfo{year}{2022}), \bibinfo{pages}{1--8}.
\newblock


\bibitem[\protect\citeauthoryear{Wu, Wu, Lyu, Huang, and Xie}{Wu
  et~al\mbox{.}}{2022b}]%
        {wu2022fedctr}
\bibfield{author}{\bibinfo{person}{Chuhan Wu}, \bibinfo{person}{Fangzhao Wu},
  \bibinfo{person}{Lingjuan Lyu}, \bibinfo{person}{Yongfeng Huang}, {and}
  \bibinfo{person}{Xing Xie}.} \bibinfo{year}{2022}\natexlab{b}.
\newblock \showarticletitle{FedCTR: Federated Native Ad CTR Prediction with
  Cross Platform User Behavior Data}.
\newblock \bibinfo{journal}{\emph{ACM Transactions on Intelligent Systems and
  Technology}} (\bibinfo{year}{2022}).
\newblock


\bibitem[\protect\citeauthoryear{Wu, Wu, Qi, Huang, and Xie}{Wu
  et~al\mbox{.}}{2022c}]%
        {wu2022fedattack}
\bibfield{author}{\bibinfo{person}{Chuhan Wu}, \bibinfo{person}{Fangzhao Wu},
  \bibinfo{person}{Tao Qi}, \bibinfo{person}{Yongfeng Huang}, {and}
  \bibinfo{person}{Xing Xie}.} \bibinfo{year}{2022}\natexlab{c}.
\newblock \showarticletitle{FedAttack: Effective and covert poisoning attack on
  federated recommendation via hard sampling}. In
  \bibinfo{booktitle}{\emph{Proceedings of ACM SIGKDD}}.
\newblock


\bibitem[\protect\citeauthoryear{Xiao, Xiao, and Eckert}{Xiao
  et~al\mbox{.}}{2012}]%
        {xiao2012adversarial}
\bibfield{author}{\bibinfo{person}{Han Xiao}, \bibinfo{person}{Huang Xiao},
  {and} \bibinfo{person}{Claudia Eckert}.} \bibinfo{year}{2012}\natexlab{}.
\newblock \showarticletitle{Adversarial label flips attack on support vector
  machines}.
\newblock In \bibinfo{booktitle}{\emph{Proceedings of ECAI}}.
  \bibinfo{publisher}{IOS Press}, \bibinfo{pages}{870--875}.
\newblock


\bibitem[\protect\citeauthoryear{Xie, Koyejo, and Gupta}{Xie
  et~al\mbox{.}}{2018}]%
        {xie2018generalized}
\bibfield{author}{\bibinfo{person}{Cong Xie}, \bibinfo{person}{Oluwasanmi
  Koyejo}, {and} \bibinfo{person}{Indranil Gupta}.}
  \bibinfo{year}{2018}\natexlab{}.
\newblock \showarticletitle{Generalized byzantine-tolerant sgd}.
\newblock \bibinfo{journal}{\emph{arXiv preprint arXiv:1802.10116}}
  (\bibinfo{year}{2018}).
\newblock


\bibitem[\protect\citeauthoryear{Yin, Chen, Kannan, and Bartlett}{Yin
  et~al\mbox{.}}{2018}]%
        {yin2018byzantine}
\bibfield{author}{\bibinfo{person}{Dong Yin}, \bibinfo{person}{Yudong Chen},
  \bibinfo{person}{Ramchandran Kannan}, {and} \bibinfo{person}{Peter
  Bartlett}.} \bibinfo{year}{2018}\natexlab{}.
\newblock \showarticletitle{Byzantine-robust distributed learning: Towards
  optimal statistical rates}. In \bibinfo{booktitle}{\emph{Proceedings of
  ICML}}. \bibinfo{pages}{5650--5659}.
\newblock


\bibitem[\protect\citeauthoryear{Zhang, Song, Gao, Chen, Bao, and Ma}{Zhang
  et~al\mbox{.}}{2019}]%
        {zhang2019your}
\bibfield{author}{\bibinfo{person}{Linfeng Zhang}, \bibinfo{person}{Jiebo
  Song}, \bibinfo{person}{Anni Gao}, \bibinfo{person}{Jingwei Chen},
  \bibinfo{person}{Chenglong Bao}, {and} \bibinfo{person}{Kaisheng Ma}.}
  \bibinfo{year}{2019}\natexlab{}.
\newblock \showarticletitle{Be your own teacher: Improve the performance of
  convolutional neural networks via self distillation}. In
  \bibinfo{booktitle}{\emph{Proceedings of ICCV}}. \bibinfo{pages}{3713--3722}.
\newblock


\bibitem[\protect\citeauthoryear{Zhuang, Wen, Zhang, Gan, Yin, Zhou, Zhang, and
  Yi}{Zhuang et~al\mbox{.}}{2020}]%
        {zhuang2020performance}
\bibfield{author}{\bibinfo{person}{Weiming Zhuang}, \bibinfo{person}{Yonggang
  Wen}, \bibinfo{person}{Xuesen Zhang}, \bibinfo{person}{Xin Gan},
  \bibinfo{person}{Daiying Yin}, \bibinfo{person}{Dongzhan Zhou},
  \bibinfo{person}{Shuai Zhang}, {and} \bibinfo{person}{Shuai Yi}.}
  \bibinfo{year}{2020}\natexlab{}.
\newblock \showarticletitle{Performance optimization of federated person
  re-identification via benchmark analysis}. In
  \bibinfo{booktitle}{\emph{Proceedings of ACM MM}}. \bibinfo{pages}{955--963}.
\newblock


\end{thebibliography}

\newpage

%\newpage
\appendix
\section{APPENDIX}
\subsection{Training Details}
The overall training process is presented in Algorithm~\ref{algo:overall}, and the \model{} is open-sourced at \url{https://github.com/deu30303/FedDefender}.

The default backbone network for our experiments is ResNet18~\cite{he2016deep}, following prior work in federated learning~\cite{li2021model,han2022fedx}.
For the attack-tolerant local meta update (Step 1), we use a first-order approximation to speed up computation. The number of nearest neighbors, k, used to perturb the label is set to 10.
In attack-tolerant global knowledge distillation (Step 2), we attach an auxiliary classifier to the output of the second residual block layer's feature map. \looseness=-1

We followed the details from original works for implementing server-side defense baselines. 
When deciding hyperparameters of Multi-Krum and Norm Bounding algorithms, we assumed that the central server already knows the upper bound of attacker numbers.
The two hyperparameters in ResidualBase algorithm, confidence interval and clipping threshold, are set to 2.0 and 0.05 respectively.

%In Scenario-1, it is assumed that the attacker has no prior knowledge of the benign clients. 
In our experiment, Scenario-2 assumes a central-server aggregation agnostic attack, where malicious users have access to benign clients' update information. We evaluate using three attack algorithms: LIE, STAT-OPT, and DYN-OPT.
For LIE, we set the intensity factor to 0.3.
For DYN-OPT, we set the initial intensity factor and threshold to 10 and $1e-5$, respectively. The optimal intensity factor value is then determined by repeatedly finding the optimal value  between the smallest and largest values, until the intensity factor change  falls under the threshold.

\begin{table}[b!]
{
\setlength{\tabcolsep}{2.5pt}
\renewcommand{\arraystretch}{1.2}
\caption{Comparison of classification accuracy with other possible baselines in Scenario-1.}
\vspace*{-0.15in}
\label{tab:recent_result}
\resizebox{1.0\linewidth}{!}{
\begin{tabular}{l|cc|cc|cc|cc}
\toprule
\multirow{2}{*}{Method} & \multicolumn{2}{c|}{CIFAR-10} &  \multicolumn{2}{c|}{CIFAR-100} & \multicolumn{2}{c|}{TinyImageNet} & \multicolumn{2}{c}{FEMNIST} \\ 
                      & ~~ Last ~~  & ~~ Best ~~  &  ~~ Last ~~ & ~~ Best ~~  & ~~ Last ~~ & ~~ Best ~~ & ~~ Last ~~ & ~~ Best ~~  \\ \midrule
No Defense            &   68.80     &  71.96   &  42.97   & 43.90  & 30.37  & 38.98 & 18.88 & 23.81  \\

+ FL-WBC         &   68.25     &  71.04   &  43.83   & 44.29  & 31.58  & 37.71 & 18.80 & 23.76\\
+ \model{}           & \textbf{78.17} & \textbf{79.96} & \textbf{51.76}  & \textbf{51.92} &  \textbf{35.59}  &  \textbf{39.68}  & \textbf{22.11} & \textbf{24.48}  \\ \hline
RFA              &  72.65      & 74.67     &  46.98  & 47.20  &  36.46 & 37.27 & 17.30 & 23.02\\
+ \model{}           	  &    \textbf{79.87}    &   \textbf{81.26}   &  \textbf{52.16}  & \textbf{52.32}  & \textbf{37.89}  &   \textbf{38.49} & \textbf{21.43} & \textbf{23.73}\\ \hline
Bucket       	  &  70.12       &  72.14    &  48.45  & 48.46  & 37.55  & 38.87 & 19.88 & 22.70 \\
+ \model{}             &  \textbf{77.57}  &   \textbf{79.37}     &   \textbf{53.32}   &  \textbf{54.44}  & \textbf{39.06}  &  \textbf{40.07} & \textbf{21.82} & \textbf{24.63} \\ \hline
\end{tabular}}
}
\vspace{-1mm}
\end{table}

\subsection{Further Results on Defense Performance}
% \cutparagraphup

% \cutparagraphup
\noindent
\textbf{Comparison with other possible defense strategy.} We have conducted additional comparison experiments with more recent baselines, including \cite{sun2021fl, karimireddybyzantine, pillutla2022robust} under a label flipping attack scenario (Scenario 1). FL-WBC is a client-side defense, while the others are server-side defenses. Based on the experimental results in Table~\ref{tab:recent_result}, we can confirm that our proposed model still offers a performance improvement for the local model when integrated with other server-side defenses. It is important to note that FL-WBC was originally designed exclusively for backdoor attacks, resulting in smaller performance gains compared with our method in untargeted attack scenarios. 

\begin{figure}[b!]
\centering
\includegraphics[width=0.7\columnwidth]{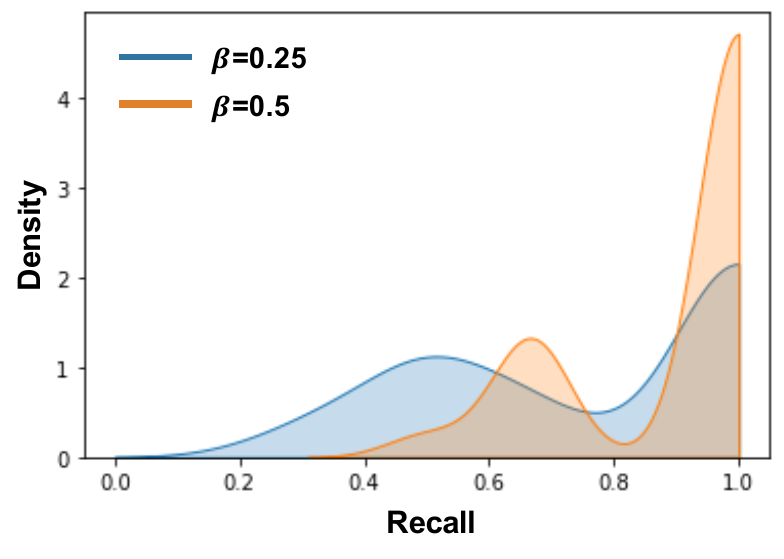}
    \vspace*{-0.12in}
    \caption{Detection recall plot of Multi-Krum with different levels of non-IID. The higher level of non-IID in data leads to the lower detection performance of the server-side defense strategy.\looseness=-1}
    \label{fig:krum_recall}
\end{figure}

\noindent
\textbf{Detection recall of Multi-Krum.} Figure~\ref{fig:krum_recall} shows the detection recall kernel density plot of the Multi-Krum algorithm, plotted against the level of non-IID data distribution. 
As non-IID data distribution becomes more extreme (i.e., lower values of $\beta$), updates from benign clients become increasingly diverse.
This makes it harder for server-side defenses to identify malicious client updates.
In this light, we propose a client-side defense strategy, \model{}, to achieve additional robustness and account for inherent model poisoning attacks during training. \smallskip

\begin{figure}[b!]
\centering
\includegraphics[width=0.7\columnwidth]{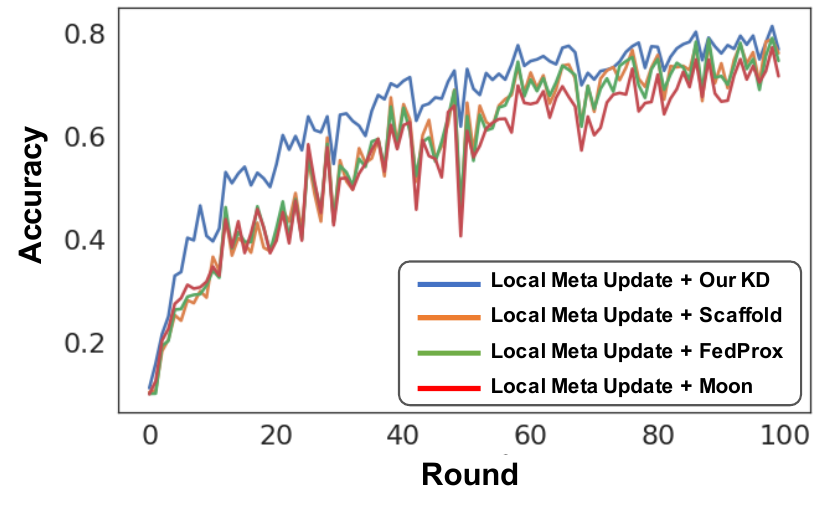}
\vspace*{-0.15in}
\caption{Performance comparison with other possible global
regularization methods across communication rounds. The proposed distillation method worked the best for all rounds.} 
    \label{fig:global_ablation}
\end{figure}

\noindent
\textbf{Comparison with other global regularization approaches across communication rounds.} 
Our proposed architecture introduces an attack-tolerant global knowledge distillation technique to mitigate the negative effects of a malicious global model during training. As an alternative to this method, one may also use existing approaches, such as FedProx~\cite{li2020fedprox}, Scaffold~\cite{karimireddy2020scaffold}, or Moon~\cite{li2021model}, in conjunction with our local meta update module.
Figure~\ref{fig:global_ablation} compares the classification accuracy of our method and alternative global regularization methods across communication rounds, without the use of any server-side defense. The models with other global regularization methods exhibit significant fluctuations in accuracy when a large number of attackers participate in each round. In contrast, our method's accuracy remains stable and consistent throughout the rounds, even when the number of participating attackers is high. 

\begin{figure*}[t!]
\centering
\includegraphics[width=1.95\columnwidth]{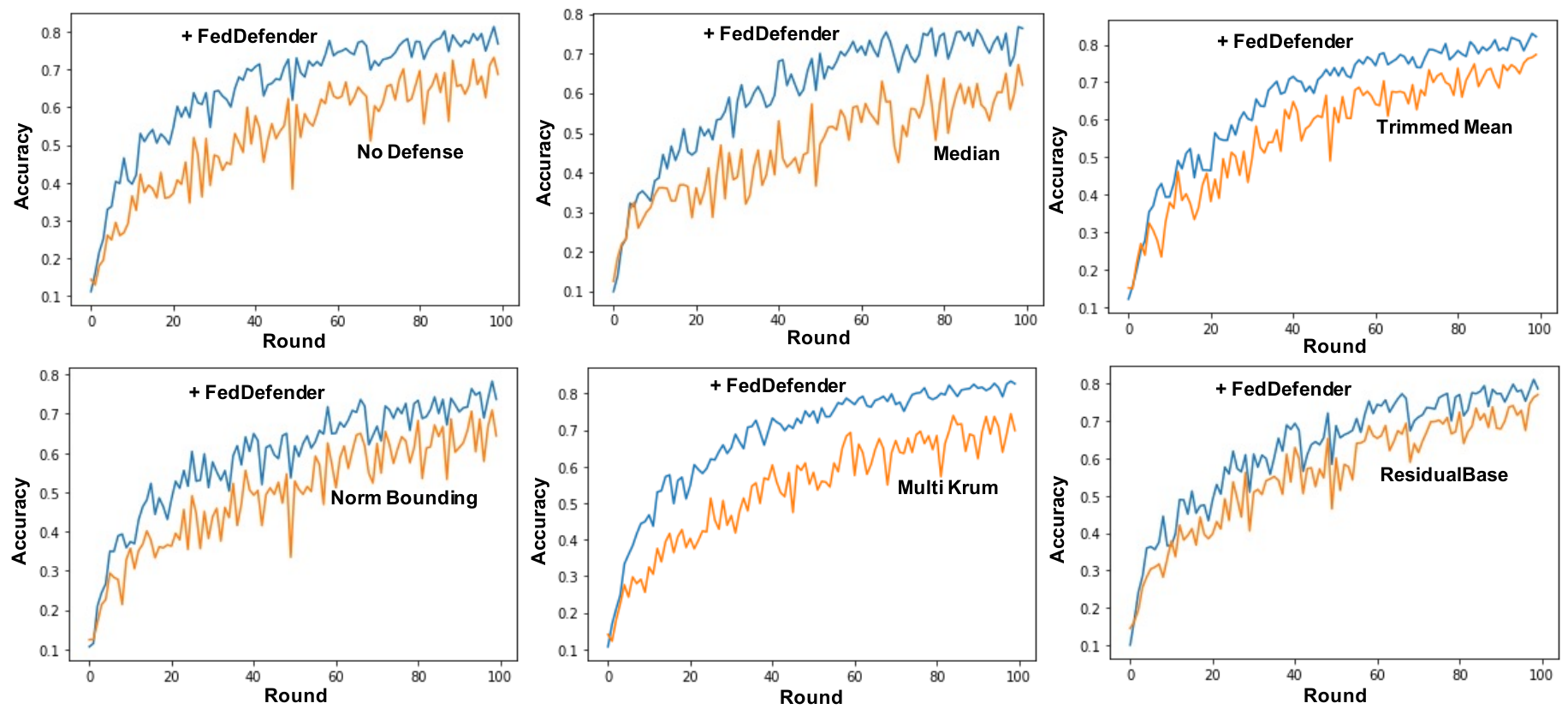}
    \cutcationup
    \caption{Performance comparison between server-side defense baselines and \model{}-enhanced versions across rounds.}
    \label{fig:training_round}
\end{figure*}

\noindent \textbf{Model comparison across communication rounds.}
The addition of \model{} results in significant improvement in robustness against model poisoning attacks, compared to relying on server-side defense only.
Figure~\ref{fig:training_round} plots the accuracy changes throughout training rounds, including six baselines: No Defense, Median, Trimmed Mean, Norm Bound, Multi-Krum, and ResidualBase.

% \cutparagraphup

\noindent
% \textbf{Model comparison via confusion matrix.}
% Figure~\ref{fig:confusion_matrix} contrasts the confusion matrices of Multi-Krum and Multi-Krum+\model{} for the CIFAR-10 dataset.
% Multi-Krum alone results in a degradation of performance across all classes.
% However, the confusion matrix for Multi-Krum+\model{} reveals a greater concentration of correct predictions along the diagonal, indicating the improved robustness brought by \model{}. \smallskip 

\smallskip

\vspace{-3mm}
\begin{algorithm*}[t!]
\DontPrintSemicolon
\SetAlgoLined
\SetNoFillComment 
\caption{Attack tolerant local update algorithm of \model{}.}
\begin{flushleft}
 \textbf{Input:} $k$-th Local Dataset $\mathcal{D}_{k}$, global model $F_\theta$, $k$-th local model $f_{\theta_{k}}$, $k$-th local auxiliary classifier model $f_{\phi_{k}}$  temperature $\tau$, learning rate $\eta$

\textbf{Output:} One epoch updated $k$-th local model $f_{\theta_k}$

\small{
\tcc{In k-th local client one epoch update process}
 
$f_{\theta_k} \leftarrow F_\theta$   \tcp*{$k$-th local model synchronization by downloading the current global model}
$F_{\theta}.\textit{detach()}$ \tcp*{Global model gradient detach for knowledge distillation}

\For{ \text{mini batch} $\mathcal{X} \in \mathcal{D}_{k}$}
{
 \tcc{\textbf{Step 1. Attack-Tolerant Local Meta Update}}

 $\mathcal{\Tilde{X}} = \{(\mathbf{x}, \mathbf{\Tilde{y}}) | (\mathbf{x}, \mathbf{y}) \in \mathcal{X} \text{ and }\mathbf{\Tilde{y}} = \text{Sample}_{\mathbf{y}}(\mathcal{N}_{k}(\mathbf{x}, \theta_k))  \}$ \tcp*{Perturbed  mini batch $\mathcal{\Tilde{X}}$ using synthetic label $\Tilde{y}$}
 $\mathcal{L}_{Perturb} = {1 \over |\mathcal{\Tilde{X}}|} \sum_{{\mathbf{x}}, {\mathbf{\Tilde{y}}} \in \mathcal{\Tilde{X}}} H(\mathbf{\Tilde{y}}, f_{\theta_k}({\mathbf{x}}))$\tcp*{Local model poisoning with synthetic noises}
$\Tilde{\theta}_k \leftarrow \theta_k - \eta \nabla_{\theta_k}  \mathcal{L}_{Perturb}$\\
$\mathcal{L}_{Meta} = {1 \over |\mathcal{X}|} \sum_{{\mathbf{x}}, {\mathbf{y}} \in \mathcal{X}} H(\mathbf{y}, f_{\Tilde{\theta}_k}({\mathbf{x}})) $\tcp*{Local model correction with meta update}
$\theta_k \leftarrow \theta_k - \eta \nabla_{\theta_k}  \mathcal{L}_{Meta}$ \\ 

\tcc{\textbf{Step 2. Attack-Tolerant Global Knowledge Distillation}}

$\mathcal{\hat{X}} = \{(\mathbf{x}, \mathbf{\hat{y}}) | (\mathbf{x}, \mathbf{y}) \in \mathcal{X} \text{ and } {\mathbf{\hat{y}}} = (1 - \alpha) \cdot \mathbf{y} + \alpha\cdot F_{\theta}(\mathbf{x}, \tau)  \}$ \tcp*{Refined mini batch $\mathcal{\hat{X}}$ using global model $F_\theta$}

$\mathcal{L}_{Global} = {1 \over |\mathcal{\hat{X}}|} \sum_{{\mathbf{x}}, {\mathbf{\hat{y}}} \in \mathcal{\hat{X}}} H(\mathbf{\hat{y}}, f_{\phi_{k}}({\mathbf{x}}))$ \tcp*{Calibrate global knowledge distillation}
$\mathcal{L}_{Self} = {1 \over |\mathcal{X}|} \sum_{{\mathbf{x}} \in \mathcal{X}} KL( f_{\theta_k}({\mathbf{x}, \tau}) || f_{\phi_k}({\mathbf{x}, \tau}))$ \tcp*{Auxiliary Self-Knowledge Distillation}
$\mathcal{L}_{KD} = \mathcal{L}_{Global} + \mathcal{L}_{Self} $ \tcp*{Calculate the entire knolwedge distillation loss}
$\mathcal{L}_{CE} = {1 \over |\mathcal{X}|} \sum_{{\mathbf{x}}, {\mathbf{y}} \in \mathcal{X}} H(\mathbf{y}, f_{\theta_k}({\mathbf{x}})) $ \tcp*{Calculate the cross entropy loss using oringal batch $\mathcal{X}$}
$\mathcal{L}_{total} = \mathcal{L}_{CE} + \mathcal{L}_{KD} $  \tcp*{Calculate the total loss}
$\theta_k \leftarrow \theta_k - \eta \nabla_{\theta_k} \mathcal{L}_{total}$ \tcp*{Update local model parameters via back-propagation}}}
\end{flushleft}
\label{algo:overall}
\end{algorithm*}

\end{document}